\documentclass[10pt, journal]{IEEEtran}

\usepackage{xcolor,soul,framed} 
\colorlet{shadecolor}{yellow}
\usepackage[pdftex]{graphicx}
\graphicspath{{../pdf/}{../jpeg/}}
\DeclareGraphicsExtensions{.pdf,.jpeg,.png}

\usepackage[cmex10]{amsmath}
\usepackage{array}
\usepackage{mdwmath}
\usepackage{mdwtab}
\usepackage{algorithm}
\usepackage{algpseudocode}
\usepackage{eqparbox}
\usepackage{url}
\usepackage{breakurl}
\usepackage{cite}
\usepackage{makecell}
\usepackage{hyperref}
\usepackage[caption=false,font=normalsize,labelfont=sf,textfont=sf]{subfig}
\usepackage{CJK}
\begin{document}
\title{Programming on Bitcoin: A Survey of Layer 1 and Layer 2 Technologies in Bitcoin Ecosystem}


\author{Guofu~Liao,~Taotao~Wang,~Qing~Yang,~Yihan~Xia,~Long Shi,~Xiang Zhao,~Xiaoxiao Wu,~Shengli Zhang,~Anthony Chan,~and~Richard Yuen

\thanks{G. Liao, T. Wang, Q. Yang, Y. Xia, X. Wu, and S. Zhang are with the College of Electronics and Information Engineering, Shenzhen University, Shenzhen 518060, China, e-mail: liaoguofu2022@email.szu.edu.cn, ttwang@szu.edu.cn, yang.qing@szu.edu.cn, xiayihan2023@email.szu.edu.cn, xxwu.eesissi@szu.edu.cn, zsl@szu.edu.cn.} 
\thanks{L. Shi is with the School of Electronic and Optical Engineering, Nanjing University of Science and Technology, Nanjing 210094, China, e-mail: slong1007@gmail.com.} 
\thanks{X. Zhao is with the College of Systems Engineering, National University of Defense Technology, Changsha 410073, China, e-mail: xiangzhao@nudt.edu.cn.}
\thanks{A. Chan and R. Yuen are with Velocity Capital, Cayman Island, e-mail: anthony@velocity.capital, richard@velocity.capital.}


}


\maketitle

\begin{abstract}
This paper surveys the innovative protocols designed to enhance the programming functionality of the Bitcoin blockchain, collectively forming a significant component of the esteemed ``Bitcoin Ecosystem''. Bitcoin relies on the Unspent Transaction Output (UTXO) model and a stack-based script language to facilitate efficient peer-to-peer payment. Nonetheless, it encounters constraints in its programming capability and throughput. The Taproot upgrade in 2021 ushered in the Schnorr signature algorithm and the P2TR transaction type to Bitcoin, thereby notably improves Bitcoin's privacy and programming capability. This upgrade catalyzes the emergence of new protocols such as Ordinals, Atomicals, and BitVM, which not only enhances Bitcoin's programming capability but also enriches the Bitcoin ecosystem. This study delves into the technical intricacies of the Taproot upgrade and investigates the Bitcoin Layer 1 protocols that harness the distinctive features of Taproot to program non-fungible tokens (NFTs) into Bitcoin transactions, including the Ordinals, Atomicals protocols, and the corresponding standards of fungible tokens constructed on Ordinals and Atomicals (i.e., BRC-20 and ARC-20). Furthermore, we categorize some selected protocols from Bitcoin ecosystem as Layer 2 solutions akin to Ethereum's Layer 2, investigating their effects on Bitcoin's performance and programming capability. By gathering and analyzing data from the Bitcoin blockchain, we obtain metrics on Bitcoin's block capacity, miner fees, and the growth trajectory of Taproot transactions initiated by the Bitcoin ecosystem protocols. The findings from data analysis affirm the beneficial impact of these protocols on Bitcoin's mainnet. This research bridges existing gaps in literature concerning Bitcoin's expanded programming capability and ecosystem protocols, offering valuable insights for both practitioners and researchers alike.

\end{abstract}


%
\IEEEpeerreviewmaketitle

\section{Introduction}

\IEEEPARstart{I}{n} 2008, Bitcoin was proposed by an author under the pseudonym Satoshi Nakamoto \cite{r10}, and successfully mined its genesis block on January 3, 2009. Bitcoin is a decentralized ledger that uses the Unspent Transaction Output (UTXO) model and scripting language to facilitate peer-to-peer money transfers, eliminating the need for intermediaries. The UTXO model effectively addresses replay attacks and speeds up Bitcoin's transaction verification \cite {r48}. In the inputs and outputs of the UTXO model, Bitcoin uses a stack-based scripting language to program transactions \cite{r7}. Over the years, this design has proven reliable, showing excellent security and stability without major vulnerabilities or issues. This success largely stems from Bitcoin's non-Turing-complete script language and its limited set of opcodes, which significantly reduce the system's attack surface. However, this design also limits Bitcoin's ability to implement complex smart contracts and advanced features, leading to reduced flexibility in programming. Bitcoin’s system also has other limitations, such as a $1$MB block size and an average block generation time of $10$ minutes, which make Bitcoin's transaction processing speed slow. Therefore, the current Bitcoin blockchain is mainly used to underpin Bitcoin as a digital asset, rather than catering to decentralized applications that demand sophisticated programming capabilities. 

In Bitcoin’s early stages, researchers and developers explored its potential applications beyond payments, especially using its transparent ledger for digital notarization services. They attempted to embed data or hashes of data into the Bitcoin ledger using the OP\_RETURN operation to extend its programming capabilities \cite {r47}. However, this method had several problems. First, the newly generated UTXOs that contain the OP\_RETURN operations could not be used or destroyed, leading to continuous growth in the UTXO set, increasing the storage burden on Bitcoin nodes. Second, to incentivize miners to include these transactions in blocks, users had to pay high transaction fees, which severely limited the widespread adoption of this application. Additionally, the programming codes embedded in the OP\_RETURN operations could only be executed and verified off-chain, requiring extra trust assumptions in off-chain nodes, thereby weakening the security and integrity of these applications.

In 2017, the Bitcoin mainnet experienced a significant technical upgrade known as Segregated Witness (SegWit) \cite{r16}. SegWit brought notable improvements by separating signature data (witness data) from the main part of the transaction and moving it to a separate section of the block. Although this change did not alter the physical block size limit (which remains $1$MB), it optimized the layout of data within blocks, effectively expanding the amount of transaction data each block could hold to approximately $4$MB, thereby increasing the throughput of Bitcoin. More importantly, the introduction of SegWit laid a solid foundation for Bitcoin's scalability. It particularly facilitated the rapid development of off-chain technologies such as the Lightning Network \cite{r12}. As a solution to the scalability of the Bitcoin mainnet, the Lightning Network enables real-time Bitcoin transactions off-chain by building a payment channel network between users. Transaction results (such as final balance updates) are submitted to the Bitcoin mainnet for confirmation only when necessary. By optimizing the data structure within blocks, SegWit directly reduces the computational and storage costs associated with creating payment channels, enabling transactions to be processed instantly, and securely closing these channels, making the Lightning Network operate more efficiently. In 2021, the Bitcoin mainnet underwent another significant upgrade--the upgrade of the Taproot proposal \cite{r17,r18,r19}. This proposal introduced the Schnorr signature algorithm and the Pay-to-Taproot (P2TR) transaction type, which not only enhanced Bitcoin’s programming capabilities, but also laid the technical possibility for implementing more complex smart contract functions. According to our data, Taproot transactions now account for more than half of all transactions on the Bitcoin mainnet (see Section \ref{sec:Data Collection and Analysis} ).

The Taproot proposal has led to the emergence of many new protocols such as Ordinals, Atomicals, BitVM, and BEVM, which aim to extend the programming capabilities of Bitcoin. And these new protocols have significantly increased activity on the Bitcoin mainnet, attracting numerous transactions and activities, and directly boosting miner revenues. This positive change has motivated miners to focus more on maintaining the security and stability of the Bitcoin mainnet, gaining widespread support from participants in the community. These new protocols can be categorized into several types: the protocols for implementing NFT functionality on the Bitcoin mainnet, such as the Ordinals Protocol and Atomicals Protocol; the standards for implementing ERC-20-like tokens, such as ARC-20, BRC-20, and Rune; and the smart contract protocols for Turing-complete programming functions, such as BitVM. Together, these protocols form a new technology and investment track known as the ``Bitcoin ecosystem'', attracting broad market interest. 

Although the above-mentioned protocols and concepts have garnered widespread attention in the industry, there is still a lack of comprehensive technical survey and academic research on these protocols. This lack of research may lead to delays in knowledge development in the relevant fields and limit the potential for innovation and application. Rencently, with respect to Taproot upgrades and technologies related to extending Bitcoin’s programming capabilities, research \cite{r4} detailed the application of ECDSA and Schnorr algorithms on the Bitcoin mainnet and discussed their implementation in scripting languages; another study \cite{r45} not only deeply analyzed the Ordinals protocol but also explored the current state and future direction of the Bitcoin ecosystem. However, there is currently a lack of systematic survey work that covers both the technologies that extend Bitcoin’s programming capabilities and the related Bitcoin ecosystem protocols. This survey paper aims to fill this gap: we not only cover the detailed descriptions and analyses of Taproot upgrades and Ordinals (as in \cite{r4} and \cite{r45}), but we also more comprehensively explain the basic elements of other relevant technologies and protocols and provide comparative analysis. Specifically, our contributions are summarized as follows: 
\begin{itemize}
\item We detail the technical intricacies of the Taproot upgrade and investigate the Bitcoin Layer 1 protocols that harness Taproot's distinctive features to program NFT into Bitcoin transactions, including the Ordinals, Atomicals protocols, and the corresponding fungible token standards constructed upon Ordinals and Atomicals (e.g., BRC-20, ARC-20).

\item Furthermore, we define some selected Bitcoin ecosystem protocols as Layer 2 solutions akin to Ethereum's Layer 2, and categorize them into four Layer 2 types. We investigate and compare the effects of these Layer 2 protocols on Bitcoin's performance and programming capability. 

\item  Finally, by gathering and analyzing data from the Bitcoin blockchain, we obtain the metrics on Bitcoin's block capacity, miner fees, and the growth trajectory of Taproot transactions initiated by the Bitcoin ecosystem protocols. And the findings from data analysis affirm the beneficial impact of these protocols on Bitcoin’s mainnet.
\end{itemize}

The structure of the remainder of this paper is as follows. Section \ref{sec:Preliminary} provides preliminary knowledge of Bitcoin. Section \ref{sec:Taproot Proposal} delves into the Taproot proposal. Section \ref{sec:Ordinals Protocol and BRC-20 Standard} examines the Ordinals protocol. Section \ref{sec:Atomicals Protocol} discusses the Atomicals protocol. Section \ref{sec:Bitcoin Layer 2 Protocols} covers Bitcoin Layer 2 protocols. Section \ref{sec:Data Collection and Analysis} introduces our data collection and analysis methods. Section \ref{conclusion} concludes the work.

\section{Preliminary}
\label{sec:Preliminary}

\subsection{UTXO Model}

\begin{figure}
  \centering
    \setlength{\abovecaptionskip}{0.cm}
  \includegraphics[width=\columnwidth]{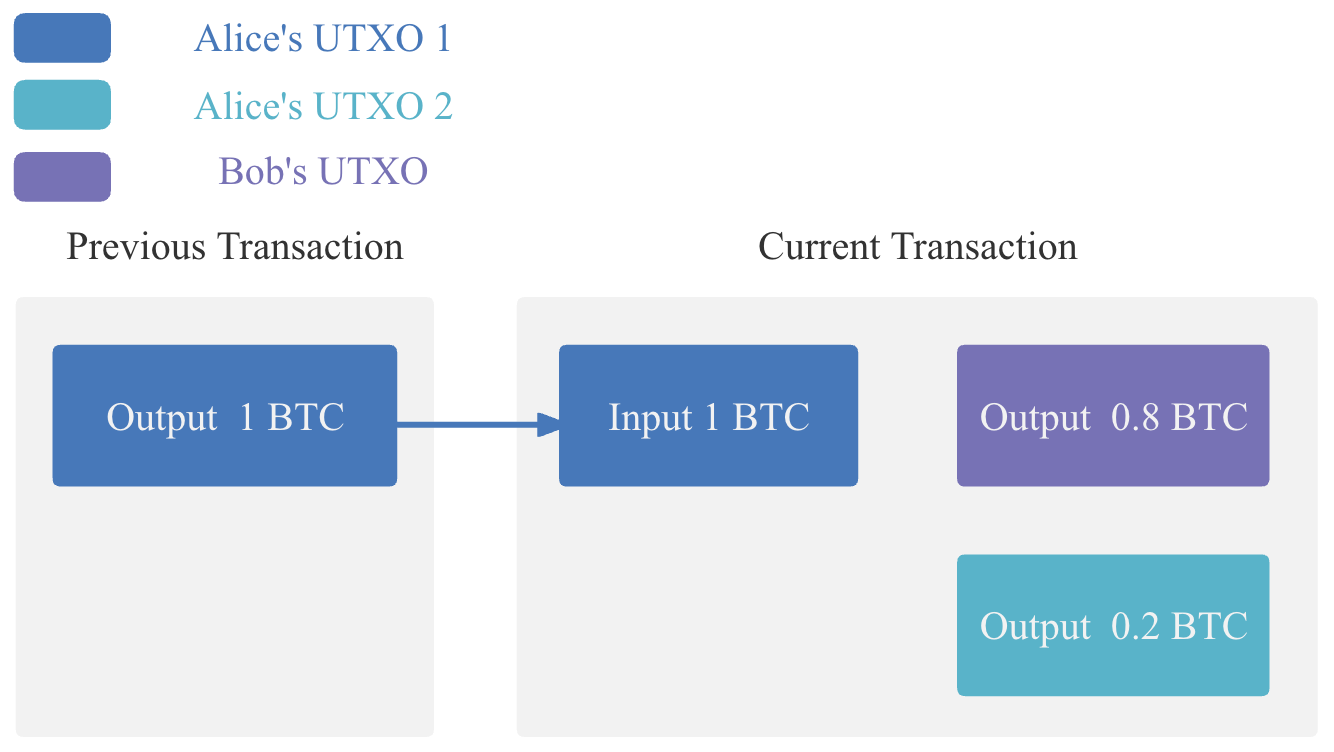}\\
  \caption{The illustration of Bitcoin's UTXO model.}\label{FIG. 01}

  \vspace{-0.2cm} \end{figure}

The addresses on the Bitcoin blockchain are computed from public keys. Each UTXO is associated with a specific owner's address and can only be used by providing a valid signature signed with the private key corresponding to the public key that generates the address. Each Bitcoin transaction consumes a number of UTXOs in its input and creates at least one new UTXOs in its output\cite{r2}\cite{r7}\cite{r11}. Although each UTXO can contain any arbitrary BTC, once created it is indivisible just like a coin that cannot be cut in half. If a UTXO is larger than the desired value of spending transaction, it still must be consumed in its entirety and must generate the change as a separate UTXO in the transaction. In the example depicted in Fig. \ref{FIG. 01}, if Alice’s address is associated with a UTXO containing 1 BTC, which is created by a previous transaction as its output and now Alice wants to pay Bob’s address 0.8 BTC, Alice will construct a transaction that takes the UTXO as input and consume the entire 1 BTC and produces two new UTXOs as output: one UTXO pays Bob’s address 0.8 BTC and the other UTXO returns 0.2 BTC as the change to Alice's own address. As a result, most Bitcoin transactions will generate change. Here, we ignore the transaction fees required to pay to miners for packing transactions onto the blockchain.

\subsection{Transaction Format}

\begin{figure}
  \centering
  \setlength{\abovecaptionskip}{0.cm}
  \includegraphics[width=\columnwidth]{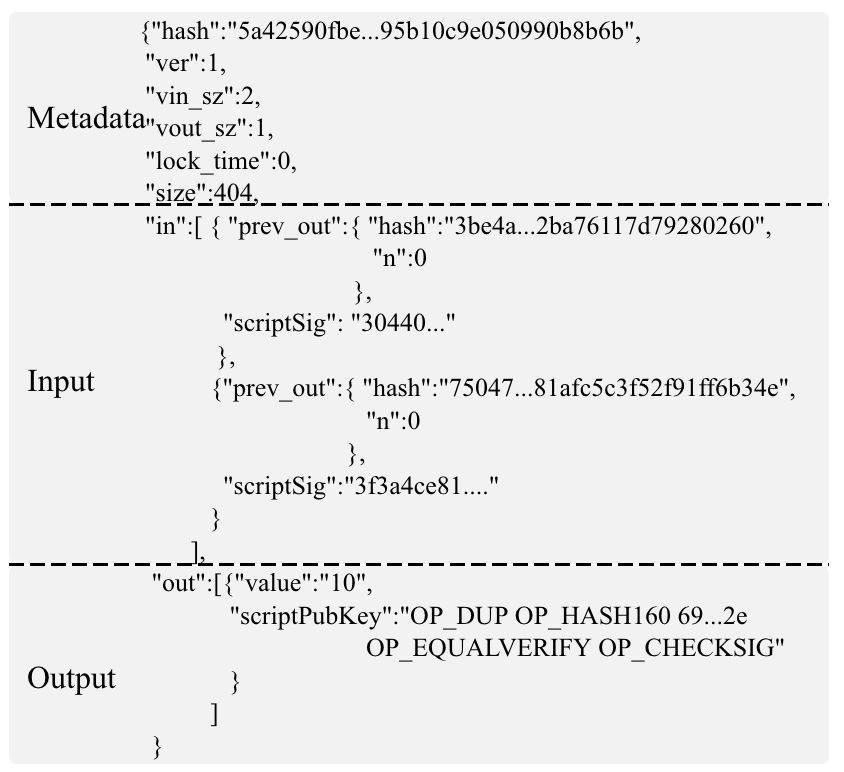}\\
  \caption{The illustration of Bitcoin's transaction format.}\label{FIG. 02}

  \vspace{-0.5cm} \end{figure}

Bitcoin’s blockchain is a series of concatenated blocks, and each of block contains a set of transactions\cite{r3}. The function of transactions is transferring BTC from a source of funds, called an input, to a destination, called an output. The format of Bitcoin transaction’s data structure is encoded into three parts: Metadata, Input, and Output, as illustrated in Fig. \ref{FIG. 02}.
The metadata part of a transaction includes the following information: the hash of the transaction, the version number of the Bitcoin software, the number of input UTXOs, the number of output UTXOs, the lock time and the memory size occupied by the transaction. The hash value is the unique identifier of each transaction, which is generated by computing the transaction content with the SHA-256 algorithm. The input part references UTXOs that are outputs from other transactions by specifying the hashes of the transaction and the sequence numbers of the outputs. To spend UTXOs, a transaction’s input also includes unlocking scripts that satisfy the spending conditions set by the referenced  UTXO. For example, the unlocking script is usually a signature proving the ownership of the Bitcoin address specifies in the locking script of the referenced UTXO. The output part creates UTXOs containing an amount of BTC, a Bitcoin address as the recipient, and the locking script that defines the condition for spending the UTXO. In most cases, the locking script will lock the output to a specific address, thereby transferring ownership of the contained BTC to the address of a new owner.

\subsection{Script Language}
The software of Bitcoin clients verifies the validity of a transaction by executing the scripts contained in the transaction, which are written in a Forth-like scripting language. Script is distinguished by its ease of use, its succinctness, and its stack-based execution engine, however, it is Turing incomplete. With fewer than 200 instructions that are known as opcodes, the scripting language of Bitcoin is designed to support many cryptographic functions\cite{r20}. Bitcoin nodes execute these script instructions to perform operations such as hashing and digital signature verification in a stack execution engine. The transaction verification within Bitcoin relies on two types of scripts: 
\begin{itemize}
    \item Locking Script: A locking script is written into a transaction output to specifies the conditions that must be satisfied to spend the output in the future. The locking script in the script program is denoted by “scriptPubKey”.

    \item Unlocking Script: An unlocking script is written into a transaction input to satisfy the conditions imposed by the locking script of the output UTXO referenced in this input. Most of the time, unlocking scripts contain a digital signature produced by the private key corresponding to the address in the locking script. The unlocking script in the script program is denoted by “scriptSig”.
\end{itemize}

When the software of Bitcoin clients verifies the validity of a transaction, the unlocking script in each input is executed along with the locking script in the output UTXO referenced by this input to see if it satisfies the spending condition. If the software of Bitcoin clients executes the scripts completely without error returning an error on top of the stack, the transaction is considered valid.  
The types of transaction outputs can be classified according to the spending conditions specified by its locking script. Important types of transaction outputs include Pay-to-PublicKeyHash (P2PKH), Pay-to-ScriptHash (P2SH), Data Recording Output (OP\_RETURN), Pay-to-Witness-PublicKey (P2WPK), and Pay-to-Taproot (P2TR). In the following, we will discuss about these types of transaction outputs with a particular focus on P2TR, which is extensively used to implement Bitcoin Layer 1 and Layer 2 protocols. 

\subsubsection{ Pay-to-PublicKeyHash (P2PKH)}

\begin{figure}

  \centering
  \setlength{\abovecaptionskip}{0.cm}
  \includegraphics[width=0.6\columnwidth]{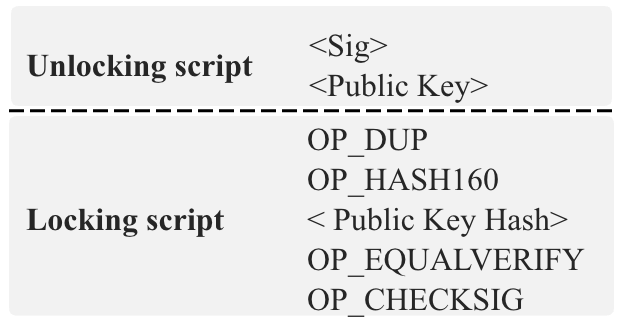}\\
  \caption{The validation script for a P2PKH transaction output.}\label{FIG. 03}

  \vspace{-0.5cm} \end{figure}

In Bitcoin, a Pay-to-PublicKeyHash (P2PKH) output of a transaction contains the hash of a public key \textless{}Public Key Hash\textgreater{} in its locking script to as the recipient of BTC in this output. To spend a P2PKH type of transaction output, it is required to supply a public key \textless{}Public Key\textgreater{} and a digital signature \textless{}Sig\textgreater{} as the unlocking script in the input that spends BTC in the P2PKH output. The two scripts (the unlocking script and the locking script) together would form the combined validation script for the P2PKH transaction output shown, as shown in Fig. \ref{FIG. 03}. The stack of Bitcoin clients will execute the combined validation script without errors if and only if the following two conditions are satisfied: 1) the public key \textless{}Public Key\textgreater{} in the unlocking script can match the hash of public key \textless{}Public Key Hash\textgreater{} in the locking script; 2) the signature \textless{}Sig\textgreater{} is generated from the private key corresponding to the public key \textless{}Public Key\textgreater{} (and thus \textless{}Sig\textgreater{} can be verified with \textless{}Public Key Hash\textgreater{}).

\subsubsection{ Pay-to-Script-Hash (P2SH)}

The Pay-to-Script-Hash (P2SH) transaction output was developed to simplify the use of more complex and functional scripts as easy as a payment script (i.e., the P2PKH script) that requires the supply of the corresponding signature to unlock. The scripts that implements complex and functional operations beyond simply paying BTC to an address (i.e., a public key hash) is called redeem scripts. An example of such redeem script is a payment script that requires the supply of multiple signatures to unlock. To enable the easy use of such redeem scripts, a P2SH output uses the hash of redeem script \textless{}Redeem Script Hash\textgreater{}, it essentially shifts the computational overhead from the transaction that contains this P2SH output to the transaction that references this P2SH output in its input.

\begin{figure*}

  \centering
  \setlength{\abovecaptionskip}{0.cm}
  \includegraphics[width=\textwidth]{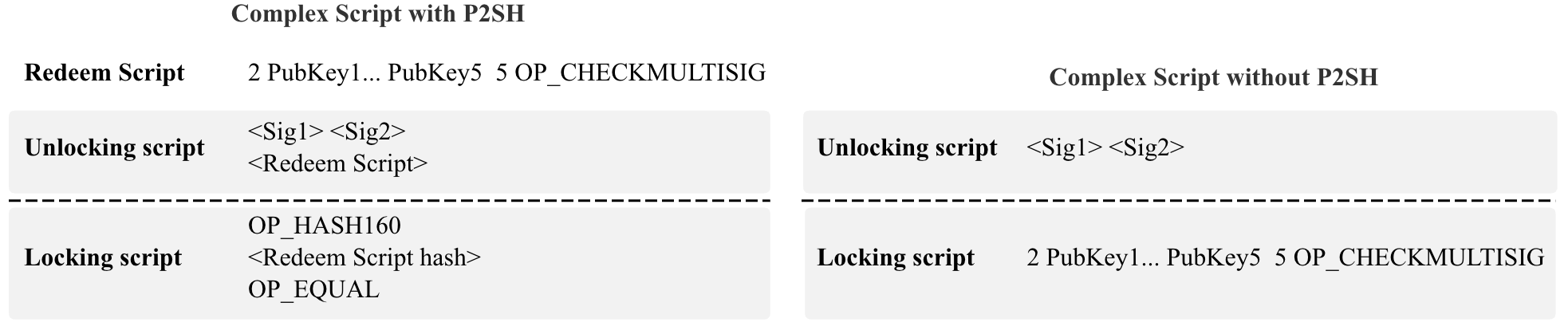}\\
  \caption{The combined validation scripts using P2SH scripts and non-P2SH scripts in a multisignature scheme.}\label{FIG. 04}

\end{figure*}

 We consider an example of redeem script that requires the supply of at least 2 signatures corresponding to the specified 5 public keys, PubKey1, …, PubKey5, to unlock the payment, as illustrated in Fig. \ref{FIG. 04}. We compare the combined validation scripts for this redeem script with and without using P2SH script. Without using P2SH script, the locking script contains the redeem script “2 PubKey1 ... PubKey5 5 OP\_CHECKMULTISIG”, indicating that two signatures are required from the five public keys to satisfy the unlock condition, and the unlocking script contains the two supplied signatures \textless{}Sig1\textgreater{} \textless{}Sig2\textgreater{}. With P2SH script, the locking script contains the hash of the redeem script \textless{} Redeem Script hash\textgreater{} and two other opcodes for verifying that the hash of the redeem script is correct when unlocking; the unlocking script contains the two supplied signatures \textless{}Sig1\textgreater{} \textless{}Sig2\textgreater{} and also the Redeem Script.  
 
Other than encapsulating redeem scripts (which are relatively large) into outputs of transactions, P2SH scripts encapsulate redeem script hashes (which are relatively small) into outputs of transactions, and thus shift the storage burden from the one  who creates the output to the one who spends the output. Therefore, with P2SH scripts, the spaces used to store the blockchain and the UTXO set can be saved.  
To spend the P2SH transaction output, the unlocking script need to contain the redeem script and the signatures that satisfies the unlocking conditions. The stack of Bitcoin client software will execute the combined validation script without errors if and only if the redeem script \textless{}Redeem Script\textgreater{} in the unlocking script can match the hash of redeem script \textless{} Redeem Script Hash\textgreater{} in the locking script.

\subsubsection{Data Recording Output (OP\_RETURN)}

\begin{figure}

  \centering
  \setlength{\abovecaptionskip}{0.cm}
  \includegraphics[width=0.6\columnwidth]{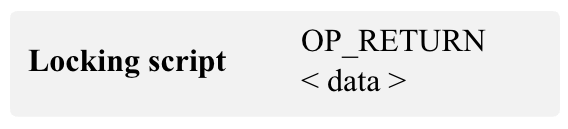}\\
  \caption{The locking script formed by OP\_RETURN.}\label{FIG. 05}

  \vspace{-0.5cm} \end{figure}

The Bitcoin Improvement Proposal 70 (BIP70) introduces a operation code that allows non-payment data to be included in transaction outputs through the use of lock scripts starting with the OP\_RETURN opcode. The OP\_RETURN operation code enables the inclusion of data unrelated to payment in the transaction output, thereby achieving the data recording function. The locking script using OP\_RETURN is illustrated in Fig. \ref{FIG. 05}, the distinguishing feature of the OP\_RETURN transaction output is that it does not need any associated unlocking script to unlock it, making it essentially an unspendable output. This feature ensures that the data recorded in the transaction output cannot be spent or modified by future transactions. Therefore, the OP\_RETURN transaction output will be excluded from the Unspent Transaction Output (UTXO) set of full node, to save the space RAM of Bitcoin full nodes.

\subsection{Digital Signature}

Digital signatures are the basic cryptographic building blocks of Bitcoin and used to verify the validity of transactions. When a user submits a transaction to the Bitcoin mainnet, the digital signature is used to satisfy the following three key functionalities: 
\begin{itemize}
    \item Authenticity: The digital signature verifies the identity of the user and the source of the funds contained in a transaction to  ensure that only the owner of the corresponding private key can authorize the spending of the associated funds. 

    \item Integrity: The digital signature is generated with recpect to the transaction data so that any changes on  the transaction data after it has been signed would invalidate the signature.  

    \item Non-repudiation: Once the digital signature is generated and it is together with the transaction recorded on the blockchain, it becomes a permanent and immutable record to ensure that the initiator cannot later refute  the transaction. 
    
\end{itemize}

The digital signature schemes employed by Bitcoin are the Elliptic Curve Digital Signature Algorithm (ECDSA) \cite{r42} and Schnorr Signature Algorithm \cite{r18}, which both utilize the principles of Elliptic Curve Cryptography (ECC). The Elliptic Curve Discrete Logarithm Problem (ECDLP) is a mathematical puzzle in ECC. The difficulty of ECDLP makes it very difficult to derive a private key from a public key, thus providing strong security guarantee for digital signatures.

ECDSA was used in the original implementation of Bitcoin clients. Later, Schnoor signature is suggested by BIP 340 and incoprtated into the implementation of Bitcoin clients after Taproot upgrade.  Utilizing Schnorr signatures in Bitcoin offers enhanced transaction efficiency, scalability, and privacy protection in comparison to ECDSA signatures. The advantages of Schnorr signatures include the following. First, Schnorr signatures allow multiple signer to cooperate to generate an aggregate signature, which can reduce the amount of signature data in a transaction when multisignature is required to unlock UTXO, and thus lower transaction fee for the transaction. Second, Schnorr signatures provide better privacy protection by aggregating multiple public keys into a single aggregated public key, hiding the each individual public keys and numbers of multisignatures in transactions. This privacy enhancement makes it difficult to distinguish multisignature transactions from ordinary single-signature transactions. In the following, we will introduce the Schnorr signature algorithm and its aggregate signature algorithm, respectively.

\subsubsection{ Schnorr Signatures}

The Schnorr signature algorithm, introduced by Claus-Peter Schnorr, is known for its simplicity, efficiency, and security \cite{r43}\cite{r4}\cite{r5}. It is widely used in various cryptographic protocols due to its compact nature and robust security. The parameters used in the Schnorr signature algorithm are defined below: 
\begin{itemize}
	\item Generator Point $G$ is a predefined point on the elliptic curve secp256k1. The generator point $G$ is a standard point chosen on the elliptic curve, which serves as a common reference for all cryptographic operations in the Schnorr signature algorithm. 
 
	\item Schnorr Private Key $sk$ is an integer selected randomly from the finite field defined by the elliptic curve secp256k1. This key is kept secret and is used for signing messages. 
 
	\item Schnorr Public Key $P$ is a point on the elliptic curve secp256k1, computed as $P=sk*G$. This key is derived from the private key and the generator point and is used for verifying signatures.
 \end{itemize}

The Schnorr signature algorithm consists of three algorithmic components:

	$(sk,P)\leftarrow KEYGEN(G)$ is the key generation algorithm that generates the Schnorr private key $sk$ and the Schnorr public key $P$ using a generator point $G$.
 
	$(sk,R)\leftarrow SIGN(k,m)$ is the signature generation algorithm that produces a signature $(s,R)$ for a given message $m$ using the private key $sk$ and the public key $P$.
 
	$1/0\leftarrow VERIFY(P,m,(s,R))$ is the signature verification algorithm that outputs a decision to accept or reject the signature $(s,R)$ using the public key $P$, the message $m$, and the signature $(s,R)$.
\begin{algorithm}[t]
    \caption{Algorithm 1: Schnorr Key Generation}\label{A1}
    
    \hspace*{0.02in} {\bf Input:} $G$ (Generator Point)\\ 
    \hspace*{0.02in} {\bf Output:} $sk$ (Schnorr Private Key), $P$ (Schnorr Public Key) 
    \begin{algorithmic}[1]
        \State Randomly select a private key $sk$ from the field defined by the elliptic curve secp256k1.
        \State Compute the public key $P$ as $P = sk \cdot G$.
    \end{algorithmic}
\end{algorithm}
 
The private key $sk$ and the public key $P$ generated by the $KEYGEN$ algorithm are foundational parameters for the Schnorr signature scheme. The $SIGN$ algorithm is executed by the signer, and the $VERIFY$ algorithm is executed by the verifier. The three algorithmic components are detailed in Algorithms \ref{A1}-\ref{A3}, respectively. Next, we introduce a particularly useful feature of Schnorr signature, i.e., the Schnorr aggregate signature.

\begin{algorithm}[t]
    \caption{Schnorr-Key Generation}\label{A1}
    
    \hspace*{0.02in} {\bf Input:} $G$ (Generator Point)\\ 
    \hspace*{0.02in} {\bf Output:} $sk$ (Schnorr Private Key), $P$ (Schnorr Public Key) 
    \begin{algorithmic}[1]
        \State Randomly select a private key $sk$ from the field defined by the elliptic curve secp256k1.
        \State Compute the public key $P$ as $P = sk \cdot G$.
    \end{algorithmic}
\end{algorithm}

\begin{algorithm}[t]
    \caption{Schnorr-Signature}\label{A2}
    
    \hspace*{0.02in} {\bf Input:} $sk$ (Schnorr Private Key), $m$ (Message)\\ 
    \hspace*{0.02in} {\bf Output:} signature $(s,R)$
    \begin{algorithmic}[1]
        \State Randomly select an integer $r$ from the field defined by the elliptic curve secp256k1. 
        \State Compute $R$ as $R = r \cdot G$. 
        \State Compute $e$ as $e = \text{hash}(R || P || m)$. 
        \State Compute the signature component $s$ as $s = r + e \cdot sk$. 
        \State Return the signature $(s, R)$, where $R$ introduces randomness and uniqueness to each signature.
    \end{algorithmic}
\end{algorithm}

\begin{algorithm}[t]
    \caption{Schnorr Verification}\label{A3}
    
    \hspace*{0.02in} {\bf Input:} $P$ (Schnorr Public Key), $m$ (Message), signature $(s, R)$\\ 
    \hspace*{0.02in} {\bf Output:} Boolean (True/False)
    \begin{algorithmic}[1]
        \State Compute $e$ as $e = \text{hash}(R || P || m)$. 
        \State Verify that $s \cdot G = R + e \cdot P$. 
        \State If the equation holds, return True; otherwise, return False.
    \end{algorithmic}
\end{algorithm}

Schnorr signatures’ mathematical foundation supports a significant advantage over ECDSA: linearity. This property allows Schnorr signatures to aggregate multiple signatures into a single signature, providing significant efficiency and flexibility benefits when developing cryptographic protocols. The parameters used in the Schnorr signature aggregate algorithm are listed as follows:
\begin{itemize}

	\item $n$: The number of participants involved in the aggregation. 
 
	\item $sk_i$: The private keys of the participants, each selected randomly from the finite field defined by the elliptic curve secp256k1. 
 
	\item $P_i$: The public keys of the participants, each computed as $P_i  =sk_i*G$, where $G$ is the generator point. 
 
	\item $P_c$: The Schnorr aggregate public key, which is the sum of the individual public keys $P_i$.
 
	\item $s_i$: The individual signature computed by each participant using their private key. 
 
	\item $S$: The aggregate signature, which is the sum of the individual signature components $s_i$.

  \end{itemize}

The Schnorr signature aggregate algorithm consists of three algorithmic components:
\begin{itemize}

	\item [1.] $(P_c) \leftarrow KEYAGG (n, sk_i, p_i, G)$ is the public key aggregation algorithm that generates the Schnorr aggregate public key $P_c$ using the private keys $sk_i$, the public key $p_i$ and the generator point $G$.
 
	\item [2.] $(S,R) \leftarrow SIGNAGG(P_i, sk_i, m, G)$ is the signature aggregation algorithm that produces an aggregate signature $(S,R)$ for a given message $m$ using the private keys $sk_i$ and the public keys $P_i$.
 
	\item [3.] $1/0 \leftarrow VERIFYAGG(P_c,m,(S,R))$ is the aggregate signature verification algorithm that outputs a decision to accept or reject the aggregate signature $(S,R)$ using the aggregate public key $P_c$, the message $m$, and the aggregate signature $(S,R)$.
 
 \end{itemize}

The three algorithmic components are detailed in Algorithms \ref{A4}-\ref{A6}, respectively.

\begin{algorithm}[t]
    \caption{ Schnorr Public Key Aggregation}\label{A4}
    
    \hspace*{0.02in} {\bf Input:} $n$ (Number of participants), $sk_i$ (Private keys), $P_i$ (Public keys), $G$ (Generator Point)\\ 
    \hspace*{0.02in} {\bf Output:} $P_c$ (Schnorr Aggregate Public Key) 
    \begin{algorithmic}[1]
        \For{each participant $i$}
            \State Randomly select a private key $sk_i$ from the field defined by the elliptic curve secp256k1.
            \State Compute the public key $P_i$ as $P_i = sk_i \cdot G$.
        \EndFor

        \State Compute the aggregate public key $P_c$ as $P_c = P_1 + P_2 + \ldots + P_n$.
    \end{algorithmic}
\end{algorithm}

\begin{algorithm}[t]
    \caption{Schnorr Signature Aggregation}\label{A5}
    
    \hspace*{0.02in} {\bf Input:} $P_i$ (Schnorr Public keys), $sk_i$ (Private keys), $m$ (Message), $G$ (Generator Point)\\ 
    \hspace*{0.02in} {\bf Output:} Aggregate signature $(S,R)$
    \begin{algorithmic}[1]
        \State Select a random number $r$.
        \State Compute $R = r \cdot G$.
        \State Compute the aggregate hash $e$ as $e = \text{hash}(m || R)$.
        \For{each participant $i$}
            \State Compute their portion of the signature $s_i$ as $s_i = k_i + e \cdot sk_i$.
        \EndFor
        \State Compute the aggregate signature $S$ as $S = s_1 + s_2 + \ldots + s_n$.
        \State Return the signature $(S,R)$.
    \end{algorithmic}
\end{algorithm}

\begin{algorithm}[t]
    \caption{Algorithm 6: Schnorr Verification}\label{A6}
    
    \hspace*{0.02in} {\bf Input:} $P_c$ (Schnorr Aggregate Public Key), $m$ (Message), signature $(S,R)$\\ 
    \hspace*{0.02in} {\bf Output:} Boolean $(\text{True/False})$
    
    \begin{algorithmic}[1]
        \State Compute $e$ as $e = \text{hash}(R || m)$. 
        \State Compute $U$ as $U = e \cdot P_c$. 
        \State Compute $V$ as $V = S \cdot G - U$. 
        \If{$V$ equals $R$}
            \State Return True (the signature is valid).
        \Else
            \State Return False (the signature is invalid).
        \EndIf
    \end{algorithmic}
\end{algorithm}

\subsection{Merkelized Abstract Syntax Trees}
BIP114 introduces Merkelized Abstract Syntax Trees (MAST), which extends Bitcoin scripts’ functionality by enabling more complex and flexible spending conditions \cite{r16}\cite{r6}. MAST combines the properties of Merkle trees and Abstract Syntax Trees (ASTs), allowing for the encapsulation and protection of Bitcoin scripts in a compressed form \cite{r41}. The proposal aims to accommodate complex spending conditions and improves privacy by hiding scripts into the leaves of MAST unexecuted branches.

\begin{figure}[t]

  \centering
  \setlength{\abovecaptionskip}{0.cm}
  \includegraphics[width=\columnwidth]{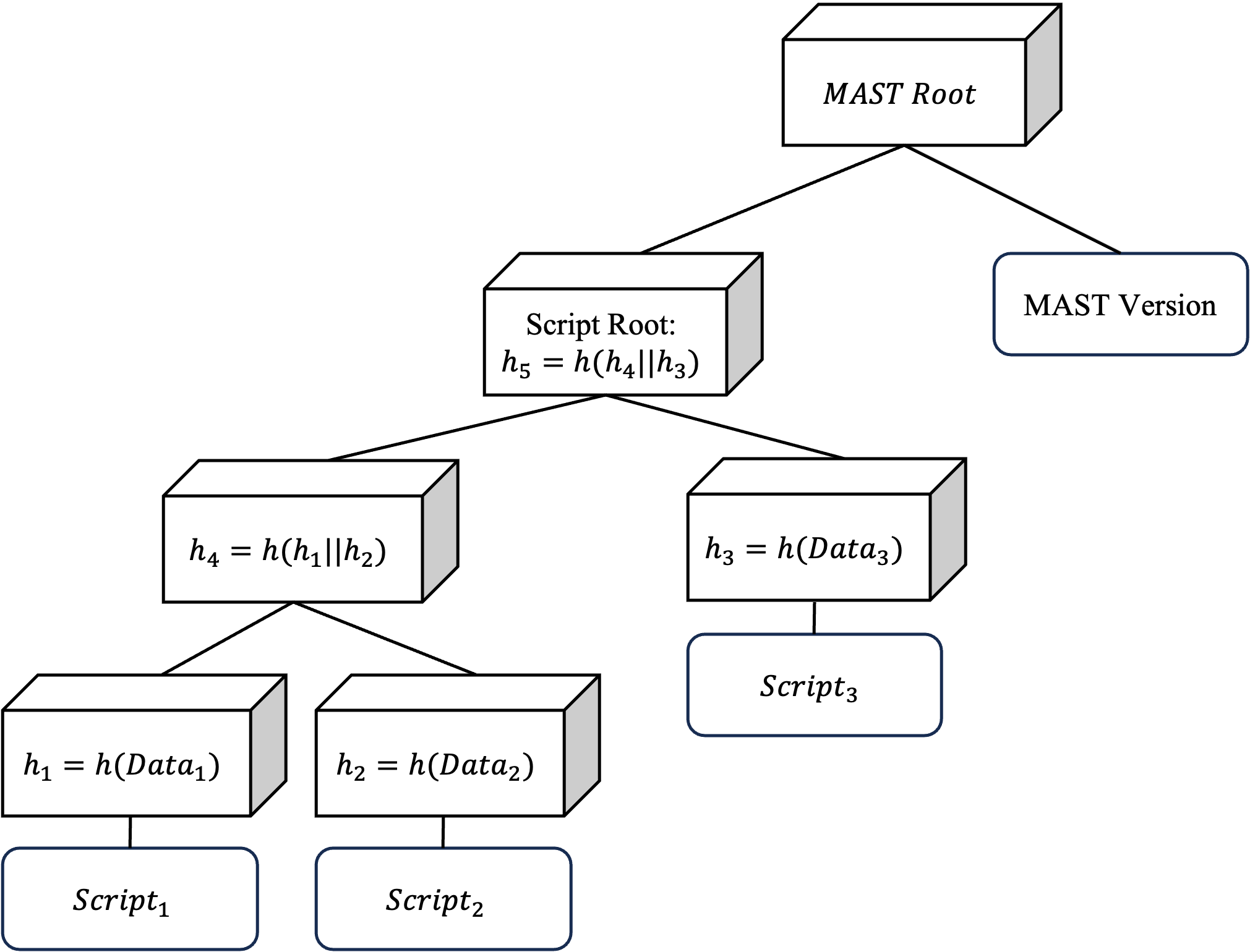}\\
  \caption{The conceptual illustration of MAST.}\label{FIG. 06}
  \vspace{-0.5cm} 
\end{figure}

As depicted in Fig. \ref{FIG. 06}, the idea of MAST is to use a Merkle tree to encode the spending conditions of a script as branches.  Each leaf node of the Merkle tree represents a spending condition, and each branch includes the path from a leaf node to the script root. For example, if a script includes multiple conditions for spending a transaction output, each condition is encapsulated in a separate branch. To spend the output, users reveal and execute only the branch with the spending condition, keeping the other branches obscured.
 
P2PKH verifies the validity of a transaction by executing the combined script (consisting of the unlocking script and the locking script) to ensure that the top of the stack returns a true at the end of execution. MAST verifies a transaction’s validity differently. To spend a transaction output that utilizes MAST, it is required to check the provided branch script and Merkle path against the Merkle root hash to ensure that the stack is empty at the end of script execution.

\section{Taproot Proposal}
\label{sec:Taproot Proposal}

In August 2017, Bitcoin implemented the upgrade of Segregated Witness, which modified the way how transaction data is stored and processed. Segregated Witness improves the transaction storage and processing efficiency by separating witness data (including signature and scripts) from the main transaction date (specifically, from the “scriptSig” field). In Segregated Witness, the witness data, which includes signatures and scripts used to unlock the outputs, is stored offchain separately from the transaction data (such as inputs, outputs, and amounts of BTC) that is recorded on the blockchain. This means that the witness data corresponding to a block does not limit by the 1MB block size. This separation allows more transactions to be included within the limited block size (1MB), and thus improving the block space utilization without increasing the block size. 

Taproot is an important upgrade to Bitcoin after Segregated Witness. The corresponding software code was merged into the Bitcoin source core repository by Pieter Wuille in October 2020. After a period of development and promotion, the proposal finally received the necessary miner support on November 14, 2021, with more than 90\% of miners reaching a consensus, and was subsequently activated in block 709,632. 

The proposal of Taproot contains three Bitcoin Improvement Proposals (BIPs), i.e., BIP340, BIP341 and BIP342. Taproot introduces a new type of transaction outputs called Pay-to-Taproot (P2TR) and the spending conditions specified by its locking script. This helps to minimize the amount of data that needs to be recorded on-chain to reveal the UTXO’s spending conditions.  

\subsection{Generate a Taproot Output Key}
In this part, we introduce the locking script of P2TR transaction output. We first introduce how to generate the Taproot output key. In a P2TR transaction output, the owner’s BTC is protected within a $33$-Bytes locking script as follows:
\begin{center}
    \textless $1$-Byte witness version\textgreater

    \textless$32$-Bytes Taproot output key $Q$\textgreater
\end{center}

The initial Byte is represented by 0x01, indicating the use of version 1 of Segregated Witness. The subsequently $32$ Bytes represents the Taproot output key denoted by $Q$. The Taproot output key $Q$ is a Schnorr aggregate public key that is computed by aggregating two Schnorr public keys as: 

\begin{center}
    $Q=P+hash(P||M)*G$
\end{center}

\noindent where $G$ represents the base point on the used elliptic curve, $P$ represents a point on the elliptic curve, and $M$ represents a Merkle tree root;  one of the aggregated two Schnorr public keys $hash(P||M)*G$, which is  derived from the MAST structure; while the other $P$ is a standard Schnorr public key, which is called internal Taproot key.
\begin{itemize}
    \item The internal Taproot key $P$ is obtained by aggregating one or multiple Schnorr public keys: $P=P_1+P_2+...+P_n$, where $P_1,  P_2,..., P_n$ respectively represents $n$ different Schnorr public keys.

    \item In MAST, the Merkle tree’s leaf nodes  contain the spending condition specified by the locking script. The Merkle tree root $M$ represents the secure commitment for the spending conditions. It compresses all spending conditions of locking script into a single hash value, thereby users can spend their UTXOs without revealing all spending conditions in the unlocking script.
\end{itemize}

The Taproot proposal integrates tagged hashing \cite{r44} to mitigate collision risks associated with hash functions. Tagged hashes are standard hashes where the tag is concatenated with the data before hashing. Taproot utilizes three distinct tagged hashes:
\begin{itemize}
    \item $Hash_{TapLeaf}$: Used to hash a leaf node.

    \item $Hash_{TapBranch}$: Used to to jointly hash sibling nodes within the Merkle tree, representing the conjoint parent node for two scripts.

    \item $Hash_{TapTweak}$: Used to jointly hash the internal public key and the root Merkle tree. 
    
\end{itemize}

\begin{figure}

  \centering
  \setlength{\abovecaptionskip}{0.cm}
  \includegraphics[width=\columnwidth]{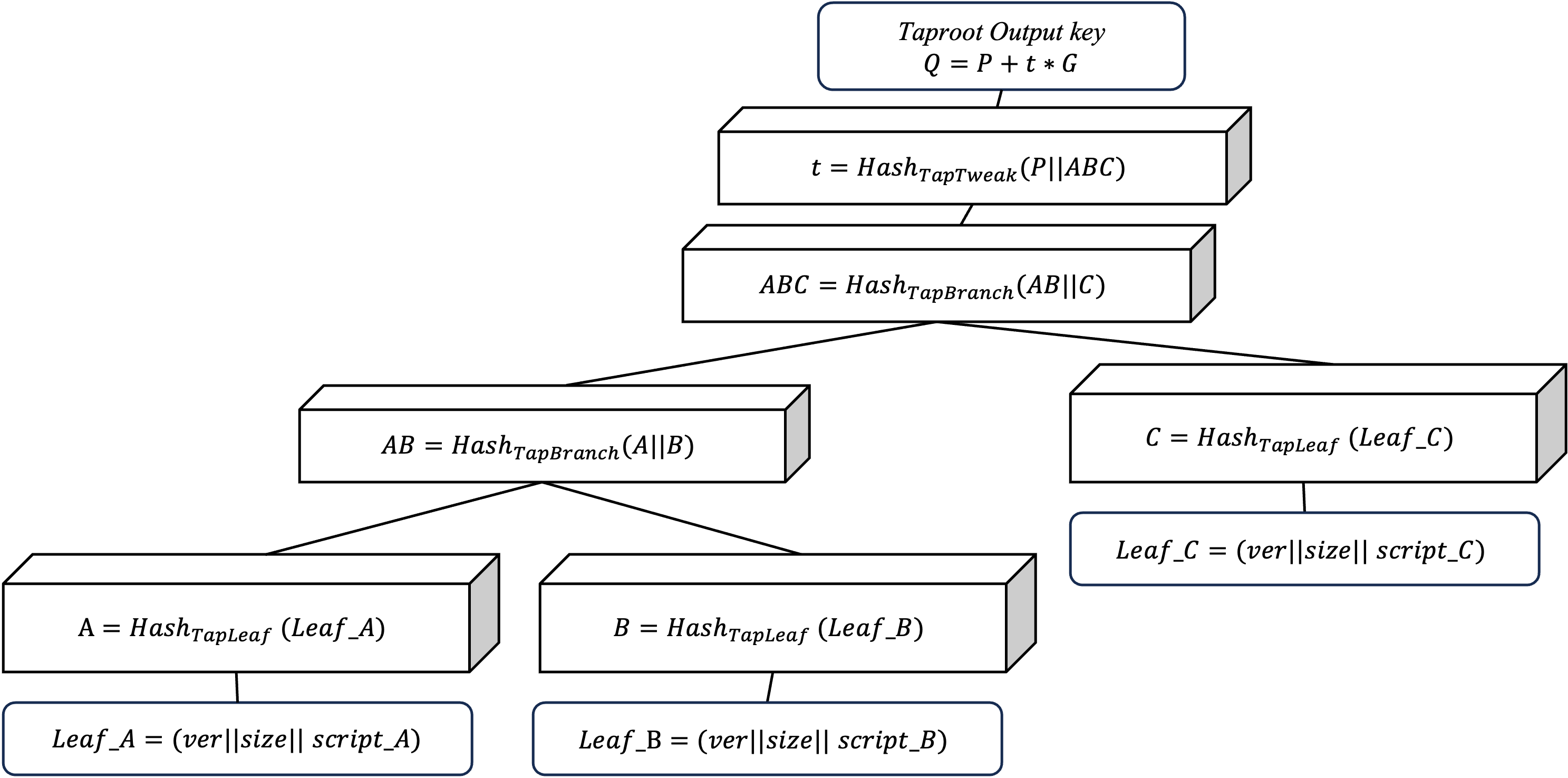}\\
  \caption{The illustration of MAST that specifies spending conditions.}\label{FIG. 07}
  \vspace{-0.5cm} \end{figure}

As depicted in Fig. \ref{FIG. 07}, the structure of the Taproot output key is generated using the MAST. The procedure begins with creating the leaf nodes that contain the spending conditions specified by the locking script. The $Hash_{TapLeaf}$ algorithm is applied to these leaf nodes to produce tags $A$, $B$, and $C$, representing the hashes of the individual spending scripts. Subsequently, the $Hash_{TapBranch}$ algorithm hashes pairs of adjacent tags, such as $A$ and $B$. This process continues recursively, combining pairs of tags until only a single tag remains. In the final phase, a tweak $t$ is generated by hashing the concatenation of the internal public key $P$ and the Merkle root $M$, using the $Hash_{TapTweak}$ algorithm.

\subsection{Types of P2TR Locking Script}
In the locking script of a P2TR transaction output, the Taproot output key $Q$ is derived by aggregating two Schnorr public keys, specifically the internal public key $P$ and another Schnorr public key $hash(P||M)*G$. In the Schnorr public key $hash(P||M)*G$, the Merkle root $M$ is computed by recursively combining the hash values of leaf nodes that contain the spending conditions specified by the locking script.  If all leaf nodes lack spending conditions, then the Merkle root $M$ is $0$. Consequently, based on the value of $M$, we can categorize the locking scripts of P2TR transaction outputs into two types:

\subsubsection{Key Path Spending $(M=0)$}

For this case, the leaf nodes do not contain spending conditions, and the Merkle root $M$ is $0$. The calculation formula for the Taproot output key $Q$ is as follows: 
\begin{center}
$Q=P+(Hash_{TapTweak} (P))*G$
\end{center}

\noindent Here, the Taproot output key $Q$ is entirely dependent on the internal public key $P$, meaning that this P2TR transaction output is governed by a single public key.

\subsubsection{Script Path Spending $(M\neq0)$}

For this case, the leaf nodes contain one or more spending conditions, and the Merkle root $M$ is computed by recursively combining the hash values of these leaf nodes. When $M\neq 0$, the calculation formula for the Taproot output key $Q$ is as follows:
\begin{center}
$Q=P+(Hash_{TapTweak} (P||M))*G$
\end{center}

The Taproot output key $Q$ is now determined by both the internal public key $P$ and the Schnorr public key $(Hash_{TapTweak} (P||M))*G$ that is computed through the Merkle root $M$.

\subsection{Taproot for Unlocking Script}
In Bitcoin, spending UTXOs requires providing spending conditions in the unlocking script so that Bitcoin client software can validate the transaction. For transaction outputs like P2PKH that are not composed of MAST, the Bitcoin client software validates the transaction output by executing the combined unlocking script and locking script in its stack and then checking if the value returned at the top of the stack is true after the script execution completes. For transaction outputs composed of MAST, the Bitcoin client software first verifies the transaction by determining if the Merkle root hash, computed from the spending conditions contained in the Merkle tree leaf node and the corresponding Merkle path provided by the unlocking script, matches the actual Merkle root hash. It then checks if the value returned at the top of the stack is empty after the script execution completes. For P2TR transaction outputs, the locking script is divided into key path spending and script path spending scripts, depending on whether the generation of the Taproot output key $Q$ involves MAST (i.e., whether the Merkle root $M$ is $0$). This means that there are two formats of unlocking scripts for spending P2TR transaction outputs. Consequently, the Bitcoin client software has two different methods to verify the validity of P2TR transaction outputs. In the following, we discuss the two types of unlocking scripts corresponding to the two types of P2TR transaction output locking scripts.   

\subsubsection{ Unlocking Script for Key path spending}

For a P2TR transaction output with a key path spending type locking script, the Taproot output key $Q$ entirely depends on the internal public key $P$, meaning that the P2TR transaction output is governed by a single public key. When spending this type of P2TR transaction output, the witness data in the unlocking script contains exactly one element, indicating a unlocking script for key path spending. In this case, the element is a signature, and it must be a valid signature for the Taproot output key $Q$. The full node verifies the transaction’s validity by executing the combined unlocking script and locking script, specifically including the execution of the Schnorr aggregate signature verification algorithm to confirm the validity of the transaction's signature.

\subsubsection{ Unlocking Script for script path spending}

For a P2TR transaction output with a script path spending type locking script, the Taproot output key $Q$ is determined by both the internal public key $P$ and the Schnorr public key $(hash(P||M)*G)$ computed from the Merkle root $M$. When spending this type of P2TR transaction output, if the witness data in the unlocking script contains three elements, it indicates an unlocking script for script path spending. The structure of a UTXO’s locking script is as shown in Fig. \ref{FIG. 07}. To spend this output, $script_A$ from the MAST leaf node is chosen to form the unlocking script, and this unlocking script is stored in the witness data:

	[Scripts]: \textless Sig\textgreater
 
	[$sript_A$]
 
	[$Control$ $block$] ( $33+32n $ Bytes)
 
Here, the [Scripts] contains a signature \textless Sig\textgreater for the transaction hash. The [$sript_A$] corresponds to the tapscript of the Merkle tree leaf node. The [$Control$ $block$] contains proof information related to $sript_A$. It includes the version number of Segregated Witness, internal public key and the proof path of $script_A$. In the following, we describe the details of the [$Control$ $block$]:

	[Tapscript Version] ($1$ Byte): $0xfe \& Control$ $block[0]$
 
	[Internal Public Key] ($32$ Bytes) 
 
	[Proof Path of $sript_A$ ]($32*2$ Bytes): $[B]$, $[C]$
 
The Bitcoin client software uses the elements in the witness data to verify the validity of the spent transaction output. This involves determining whether the Merkle root hash calculated from the leaf node and Merkle path provided by the unlocking script matches the actual Merkle root hash.

\subsection{Application of Taproot in Multisignature}
In P2SH transaction outputs, when spending conditions require multiple signatures, the size of the unlocking script (i.e., witness data) increases with the number of signers. This is because the locking script only contains the redeem script hash \textless Redeem Script Hash\textgreater, while multiple signatures and the redeem script \textless Redeem Script\textgreater are included in the unlocking script. In fact, a P2SH redeem script can contain up to 15 public keys for multisignature verification. This limit is based on the following reasons:

1) The OP\_CHECKMULTISIG opcode is designed to verify up to $15$ public keys. This design choice balances the efficiency and reliability of script execution with the need to support a reasonable number of signatures in multisignature transactions.

2) Bitcoin has a limit on the total size of each transaction, with a maximum of $1$MB. Considering that each public key occupies about $33$ Bytes and each signature about $72$ Bytes, an unlocking script with fifteen signatures and corresponding public keys approaches this maximum limit.

In contrast, using the Schnorr signature algorithm in P2TR transaction outputs can achieve multisignature schemes with more public keys for verification\cite{r26}\cite{r34}. The Schnorr signature algorithm leverages its linear property to aggregate multiple participants’ public keys $pk_i,i=1,2,...,n$ into a single public key $pk$, and aggregate each participant’s signature $s_i,i=1,2,...,n$ into a single signature $S$. This aggregated public key $pk$ is then used to verify the signature $S$. This feature allows locking scripts to include more signatures, creating more complex spending conditions. Next, we discuss the $n-$of$-n$ and $m-$of$-n$ multisignature schemes using the Schnorr signature algorithm.

\subsubsection{ $n-$of$-n$ Signature Scheme}

In an $n-$of$-n$ signature scheme, each participant provides a valid signature to satisfy the spending condition of the transaction output. This means all participants first generate their own Schnorr public key $pk_i,i=1,2,...n$, and use the linear combination property of the Schnorr signature algorithm to aggregate these public keys into a single aggregate public key $pk$, generating a Taproot output key $Q$ as part of the locking script. To satisfy the spending conditions specified by the locking script and construct the unlocking script, all participants generate independent Schnorr signatures $s_i$ on the transaction hash and then aggregate these signatures into a single aggregate signature $S$ as part of the unlocking script. Bitcoin client software executes the combined script (consisting of the unlocking script and locking script) to verify the aggregate signature. If the top of the stack is true at the end of execution, the transaction output can be spent.

\subsubsection{ $m-$of$-n$ Signature Scheme}

\begin{figure}

  \centering
  \setlength{\abovecaptionskip}{0.cm}
  \includegraphics[width=\columnwidth]{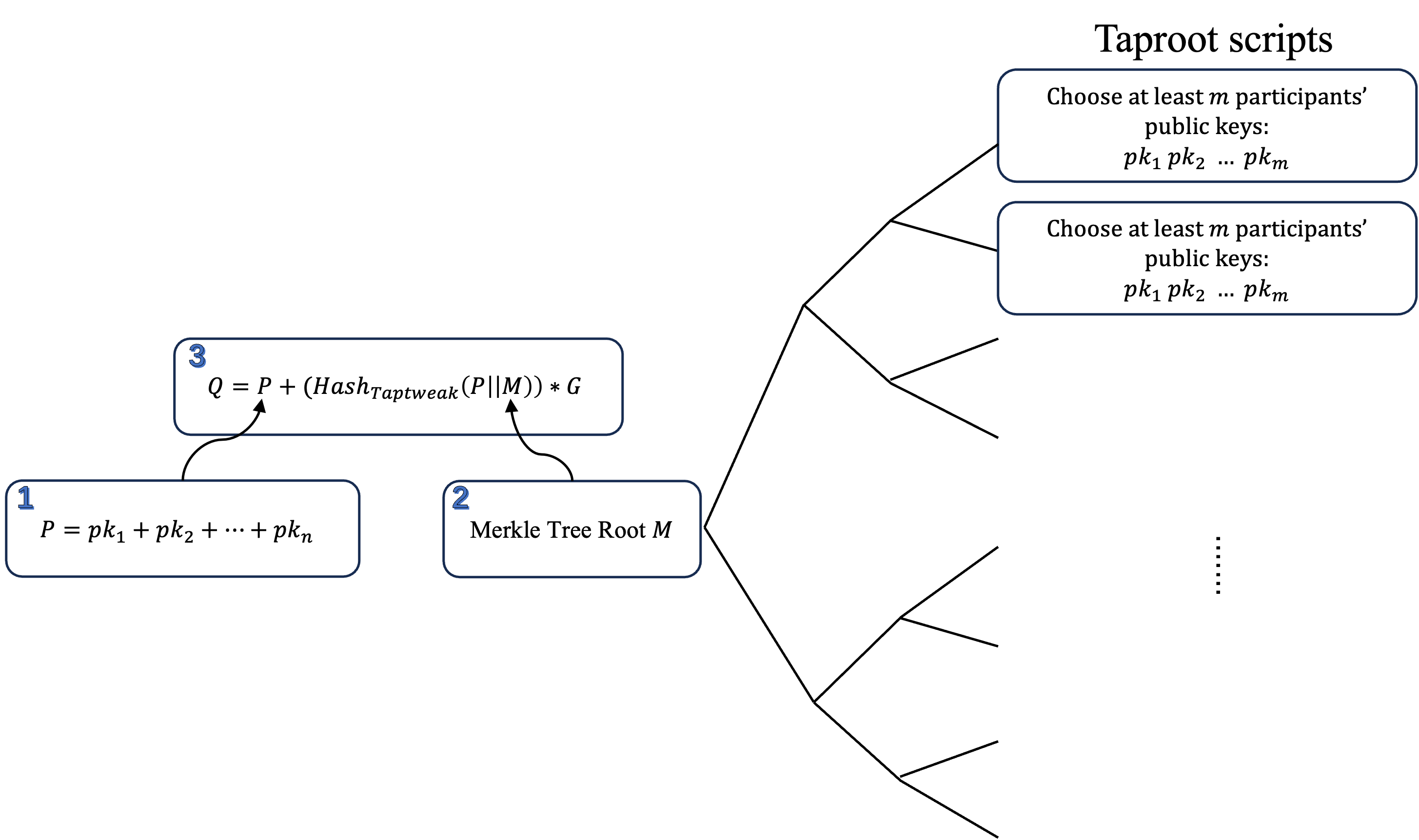}\\
  \caption{The process of encoding spending conditions with an $m$-of-$n$ multisignature scheme in a P2TR transaction output.}\label{FIG. 08}
  \vspace{-0.5cm} \end{figure}

In an $m-$of$-n$ signature scheme, at least $m$ out of $n$ participants (where $m < n$) must provide valid signatures to satisfy the spending conditions of the transaction output. The following discusses the process of implementing the $m-$of$-n$ multisignature scheme in P2TR transaction outputs:

\begin{itemize}

    \item  {\bfseries Spending Condition Encoding Process}: As depicted in Fig. \ref{FIG. 08}, the process of generating the locking script of P2TR transaction output is demonstrated. First, each participant generates their own private key and corresponding public key. These public keys are aggregated into a single aggregate public key, called the internal public key $P$. Subsequently, the $m$-of-$n$ multisignature scheme (where the spending condition in each leaf node requires at least $m$ out of $n$ participant public keys) is encoded into the Taproot script as leaf nodes of the Merkle tree to generate the Merkle tree root. An additional Schnorr public key $hash_{TapTweak}(P||M)*G$ is computed. Finally, the internal public key $P$ is combined with $hash_{TapTweak}(P||M)*G$ to generate the Taproot output key $Q$.

    \item {\bfseries Spending Process}: To satisfy the spending condition of the transaction output, at least $m$ participants sign the transaction hash using their private keys $sk_i$. These signatures are aggregated into a single signature $S$ using the Schnorr signature algorithm as the first element of the witness data. Next, the public keys $pk_i$ of the m participants are provided to generate the Taproot script corresponding to the Merkle tree leaf node as the second element of the witness data. The Merkle path from the leaf node to the Merkle root is provided as the third element of the witness data. Finally, the Bitcoin client software uses the provided aggregate signature, leaf node script, and Merkle path proof to verify the validity of the transaction and then decides whether to execute the spend.
    
\end{itemize}

\section{Ordinals Protocol and BRC-20 Standard}
\label{sec:Ordinals Protocol and BRC-20 Standard}

The Taproot upgrade introduced new solutions for implementing more complex programming functions on the Bitcoin blockchain, particularly showing potential in emerging fields such as non-fungible tokens (NFTs) and decentralized finance (DeFi). Based on the technical advantages brought by the Taproot upgrade, Casey Rodarmor launched the Ordinals protocol in January 2023, aiming to enable NFT functionality on Bitcoin\cite{r22}. This section will introduce the Ordinals protocol and its derivative BRC-20 token standard.

\subsection{Ordinals Protocol}
The Ordinals protocol primarily introduces a novel precise numbering system for Bitcoin's smallest units (i.e., “sat” or satoshi), allowing each sat to carry additional metadata (such as text, images, code, etc.), thereby enabling support for non-fungible tokens. This numbering method not only enhances Bitcoin's programmability but also provides potential for developing complex smart contracts and decentralized applications on the Bitcoin mainnet. The core of the Ordinals protocol includes three parts: Number Scheme, Transfer Scheme, and Inscriptions. The following sections will provide a detailed introduction to each part. 

\subsubsection{ Number Scheme}

The Bitcoin protocol sets a series of periodic events during its block generation process, such as the Bitcoin halving every 210,000 blocks, the adjustment of block difficulty every 2016 blocks based on the hash rate, etc. The most frequent periodic event is the generation of a new block every 10 minutes. Additionally, here is a difficulty adjustment every two weeks based on the change in hash rate, and the halving event every four years, which significantly impacts Bitcoin’s supply and inflation rate. Furthermore, approximately every 24 years (i.e., after six halvings), halving and difficulty adjustment occur simultaneously, forming a conjunction point. The first conjunction point is expected to appear around 2032.

The Ordinals protocol considers that new sats minted in blocks produced at the above periodic events have some rarity. Depending on the block’s periodic events and the sequence of new sats within the block, new sats have different rarity levels and can be classified as Common, Uncommon, Rare, Epic, Legendary, and Mythic. For example, the first sat appearing in the genesis block is within the first period and belongs to the first block of both the halving and difficulty adjustment periods, making it of Legendary grade.

The Ordinals protocol provides various methods for representing and classifying sats. Here are examples of different representation methods:
\begin{itemize}

\item Integer representation: $1947673583222332$, using numerical values to represent a specific sat, corresponding to its mining sequence number.

\item Decimal representation: $806277.458222332$, using two numbers to represent a specific sat, the first indicating the block height and the second indicating the offset within the block.

\item Degree representation: $0°176277'1893''458222332'''$, with a detailed structure as follows: $A°B'C''D'''$:

$A$: Period number, starting from $0$.

$B$: Block index within the halving period.

$C$: Block index within the difficulty adjustment period.

$D$: Sat index within the block.

\item Percentile representation: $92.74636120784638\%$, indicating a sat’s position within the overall Bitcoin supply.

\item Name representation: abakjcudpll, using alphabetic characters to assign a unique name to the sat.
\end{itemize}

According to the degree representation method, the rarity classification of sats is as follows: any sat with a non-zero value in any field is considered common, a sat with field $D$ as zero is uncommon, a sat with fields $C$ and $D$ as zero is rare, a sat with fields $B$ and $D$ as zero is epic, and a sat with all fields as zero is mythic. Within the Ordinals protocol framework, the distribution of rare sats is approximately: 2.1 quadrillion common sats, about $6,929,999$ uncommon sats, $3,437$ rare sats, $32$ epic sats, $5$ legendary sats, and only one mythic sat.

\subsubsection{ Transfer Scheme}

In the Ordinals protocol, sats are assigned serial numbers based on the sequence in which new BTC is mined, with the first sat in the first block assigned serial number $0$, and subsequent sats incrementing sequentially. However, since multiple sats can be included in one transaction output, if it is necessary to accurately identify and transfer specific sats (e.g., those with higher rarity) in a transaction, it becomes complex. To address this issue, Ordinals introduces a transfer scheme to determine how sats are transferred from transaction inputs to transaction outputs. 
This scheme uses an algorithm to ensure that sats in the transaction input are transferred to the sats in the transaction output in the order of their serial numbers in a first-in, first-out (FIFO) manner, constrained by the size and sequence of the transaction inputs and outputs. This transfer scheme ensures the uniqueness and traceability of each sat, allowing precise management and tracking of specific sats even in complex transactions. It solves the problem of rearranging and reallocating sats during transactions, ensuring reliability and accuracy when using the Ordinals protocol to implement NFTs and other advanced applications on the Bitcoin mainnet. 

We use the example in Fig. \ref{FIG. 09} to illustrate the sat transfer scheme in the Ordinals protocol in more detail. Suppose address A initiates a transaction to transfer 20 BTC to address B. In this example, address A, with 50 BTC input, contains sats ranging from [0 to 4,999,999,999]. The transaction transfers 30 BTC back to address A, containing sats ranging from [0 to 2,999,999,999]; the remaining 20 BTC is transferred to address B, containing sats ranging from [3,000,000,000 to 4,999,999,999].
\begin{figure}
\label{FIG. 09}

  \centering
  \setlength{\abovecaptionskip}{0.cm}
  \includegraphics[width=0.8\columnwidth]{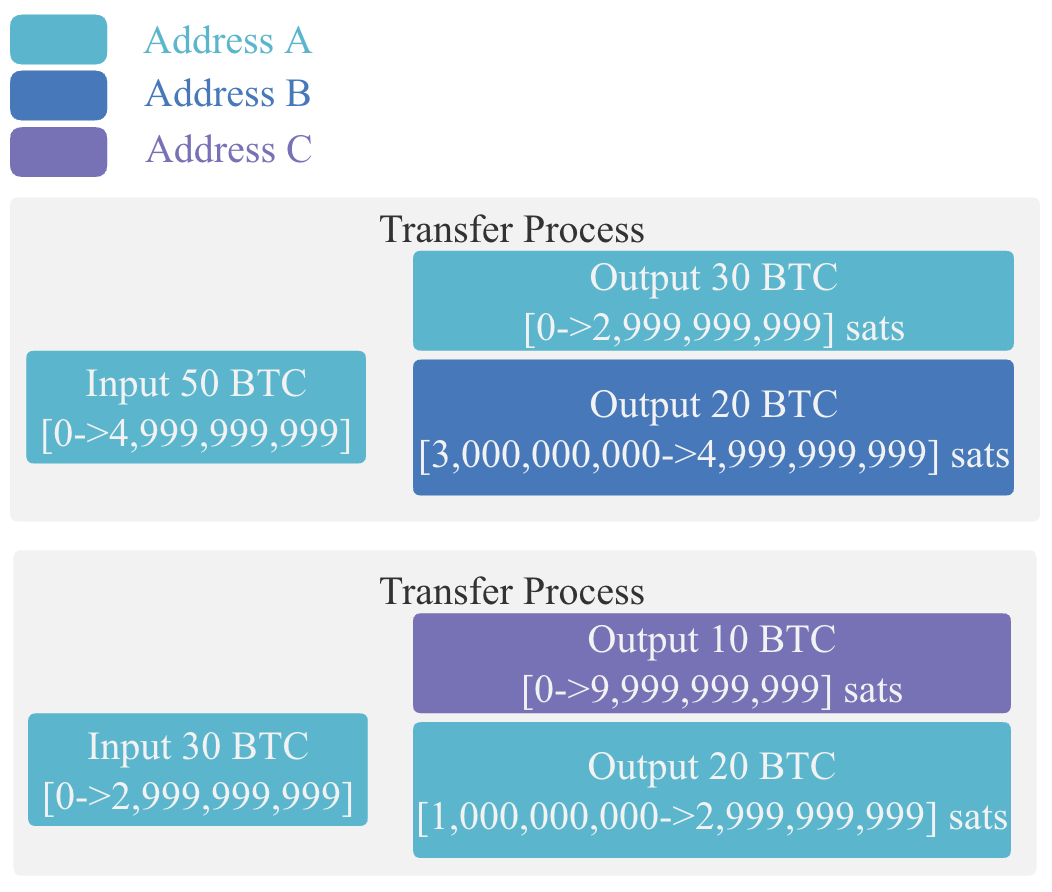}\\
  \caption{An illustrating example of the sat transfer process in the
Ordinals protocol.}\label{FIG. 09}
  \vspace{-0.5cm} \end{figure}

Then, if address A wants to transfer 10 BTC to address c while retaining the 2,599,999,999th sat (as the sat might be very rare), the owner of address A needs to control the sequence of the transaction outputs to ensure that the sat remains in address A’s UTXO set, as shown in Fig. \ref{FIG. 09}:
\begin{itemize}

\item Address A with 30 BTC input contains sats ranging from [0 to 2,999,999,999].

\item The first output transfers 10 BTC to address C, containing sats ranging from [0 to 999,999,999].

\item The second output transfers the remaining 20 BTC back to address A, containing sats ranging from [1,000,000,000 to 2,999,999,999].

\end{itemize}

In this way, the transfer scheme in the Ordinals protocol ensures the management and precise tracking of specific sats, maintaining reliability and accuracy even in complex transaction scenarios.

\subsubsection{ Inscriptions Process}

\begin{figure}

  \centering
  \setlength{\abovecaptionskip}{0.cm}
  \includegraphics[width=\columnwidth]{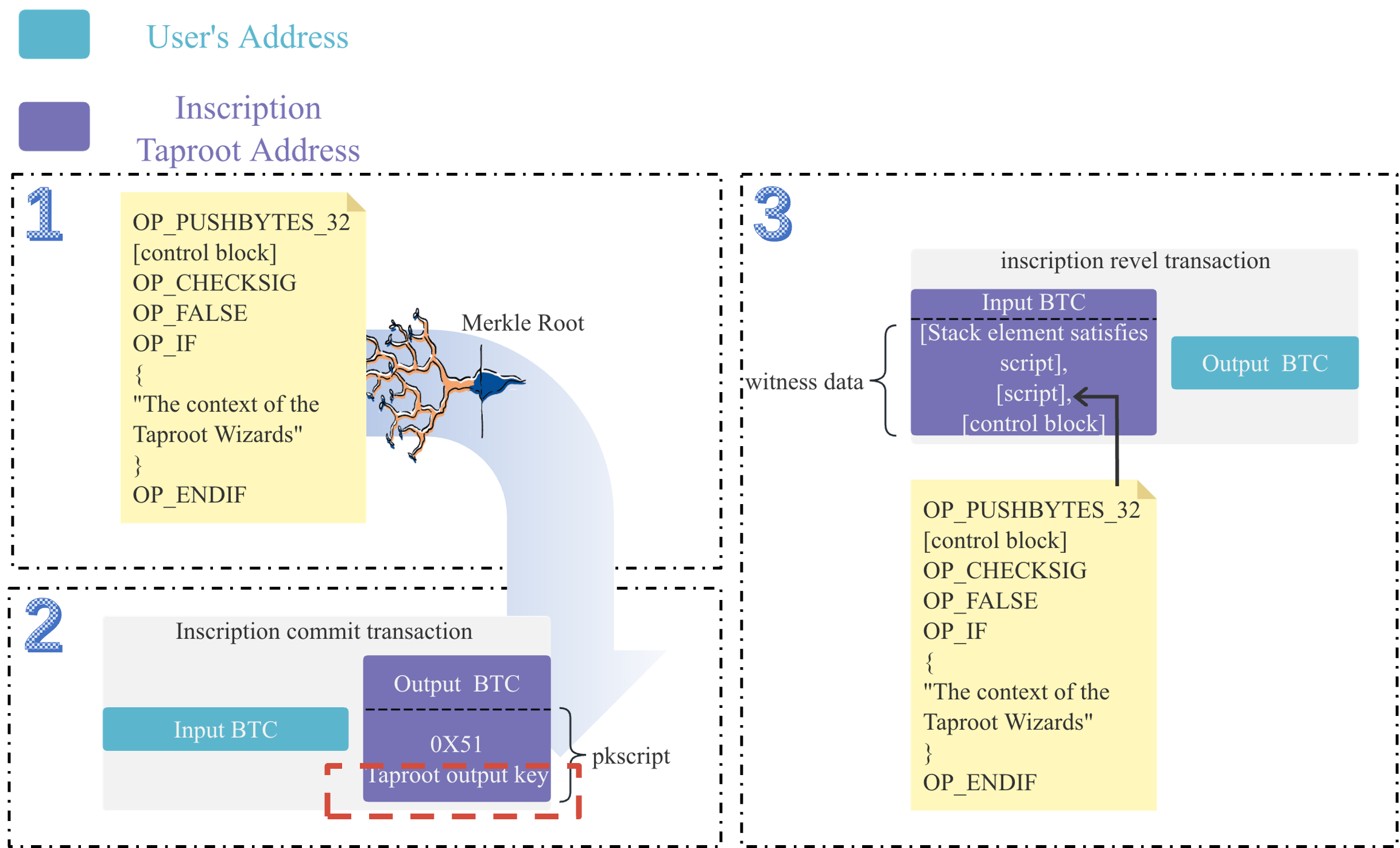}\\
  \caption{The inscription process of Taproot Wizards.}\label{FIG. 10}

  \vspace{-0.5cm} \end{figure}

The Ordinals protocol introduces a mechanism called inscriptions, allowing any content to be “inscribed” onto sats, thereby creating native digital artifacts on Bitcoin, commonly known as NFTs. The inscription process mainly involves two transactions, called the commit transaction and the reveal transaction. First, a Taproot output key is generated in the commit transaction, which is essentially a commit to the inscription content script. After completing the commit transaction, the reveal transaction is executed. This step spends the output of the commit transaction, storing the inscription content on the blockchain for anyone to verify. Fig. \ref{FIG. 10} illustrates the steps of the inscription process using the text string “Taproot Wizards” as the inscription content. Below is a description of the steps shown in the figure.

In the commit transaction, the locking script of its transaction output contains two fields: [taproot output key] and [0x51]. The [taproot output key] field is a Schnorr aggregate public key, obtained by combining the internal Taproot key $P$ and the hash value of the Merkle root, then multiplying it by the elliptic curve generator $G$, and finally aggregating with the internal Taproot key $P$. In the leaf node of the Merkle tree, the inscription content “Taproot Wizards” is used as the spending condition. The [0x51] field indicates version $1$ of Segregated Witness. In the commit transaction, the user submits the inscription content to the Bitcoin blockchain in the form of the Taproot output key, but the content is not yet publicly available on the Bitcoin mainnet. Subsequently, the reveal transaction is initiated, requiring the unlocking script to provide the spending conditions for the UTXO generated in the commit transaction. In the Taproot protocol, the spending conditions are stored in the witness data, containing three elements: {[Scripts], [Tapscript], [control block]}. The Bitcoin client software uses the elements in the witness data to verify the validity of the spent transaction output. First, the Bitcoin client software computes the Merkle root hash from the leaf node and Merkle path provided by the unlocking script and checks if it matches the actual Merkle root hash. If the root hash matches, it executes stack operations with the [Scripts] from the witness data and the combined script of the locking script. If the top stack element is empty at the end, the reveal transaction is valid. At this point, the inscription content “Taproot Wizards” is publicly available on the Bitcoin mainnet in the form of Tapscript in the unlocking script.

Through the inscription mechanism, text and image content can be inscribed onto specific sats. Once the inscription content is inscribed onto a sat and published on the Bitcoin blockchain, it will be permanently recorded on the blockchain, issuing an NFT that cannot be tampered with or deleted. Transferring the inscribed sat during Bitcoin transactions is equivalent to transferring ownership of the inscribed content, thus achieving the function of NFT ownership transfer.

\subsection{BRC-20}

\begin{figure*}

  \begin{center}  
  \includegraphics[width=\textwidth]{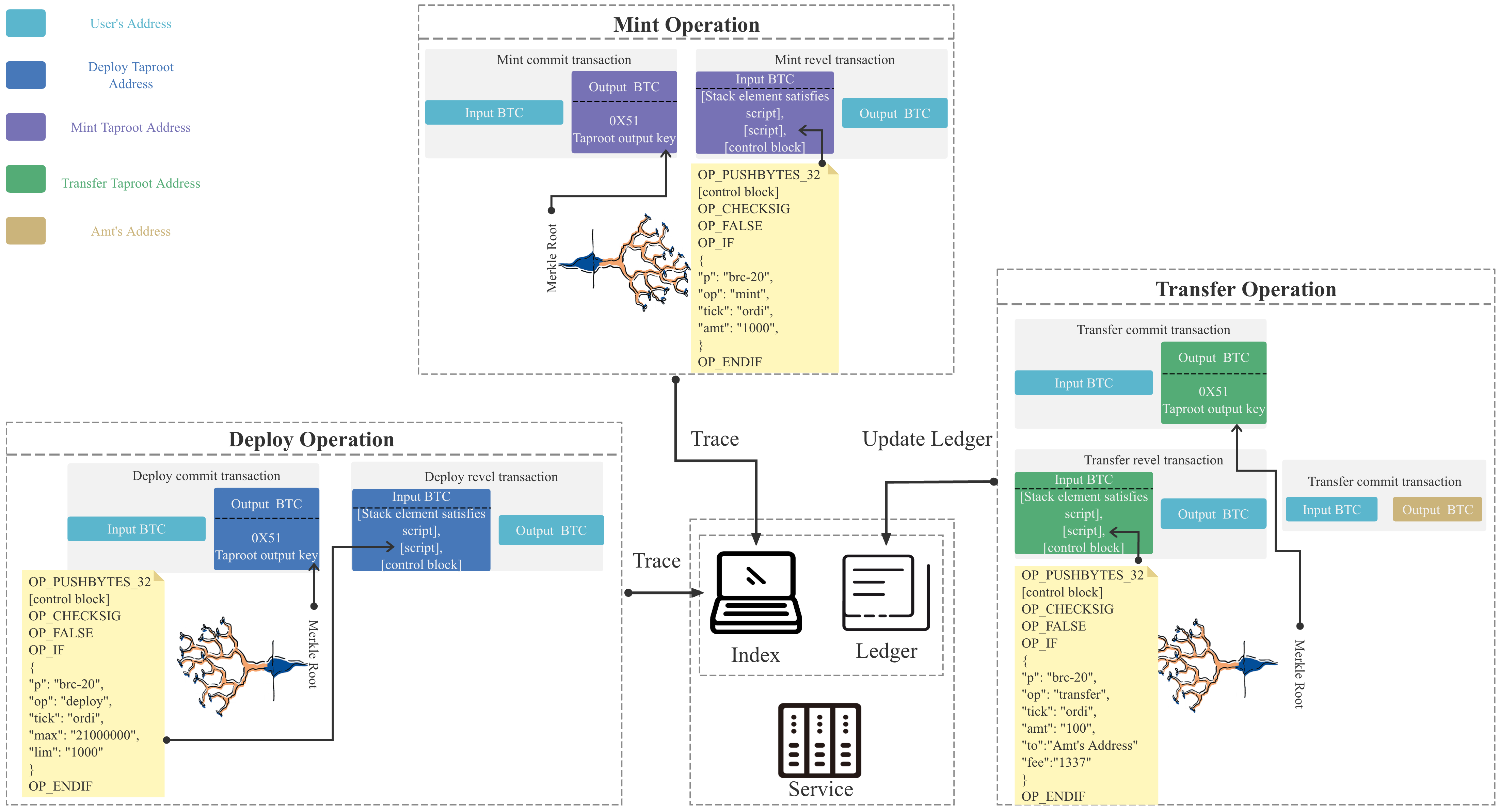}\\
  \caption{The deploy, mint, and transfer processes of BRC-20.}\label{FIG. 11}
  \end{center}
  \vspace{-0.5cm} 
\end{figure*}

The BRC-20 token standard, leveraging the inscription mechanism of the Ordinals protocol, enables the creation of fungible tokens on Bitcoin’s base layer. The name “BRC-20” is inspired by the Ethereum ERC-20 standard. BRC-20 defines a standardized method for deploying, minting, and transfer tokenized assets on the Bitcoin blockchain by inscribing specific JSON-formatted data onto sats. It is important to note that the BRC-20 token standard requires additional third-party ledgers, known as “indexers,” to track users’ token balances. Since BRC-20 is a token standard that cannot implement smart contract functionality, miners on the Bitcoin blockchain will not recognize operations such as the deploy, mint, or transfer of tokens inscribed on sats. Indexers are needed to ensure the new fungible tokens comply with the BRC-20 standard. 
To illustrate the implementation of BRC-20, this section uses the issuance of a BRC-20 token named “ordi” as an example. Fig. \ref{FIG. 11} demonstrates the operational processes related to “ordi” BRC-20 tokens, describing their operation of deploy, mint, and transfer on Bitcoin.

The operations related to fungible tokens include: Deploy, Mint, and Transfer.

\subsubsection{The Operation of Deploy}

For tokens that have not been deployed, users can deploy token information to Bitcoin through commit and reveal transactions. The JSON format of a deploy transaction is as follows:

\hspace{0.8cm}\{“p”: “brc-20”, 

\hspace{0.8cm} “op”: “deploy”, 

\hspace{0.8cm} “tick”: “ordi”,

\hspace{0.8cm} “max”: “21000000”, 

\hspace{0.8cm} “lim”: “1000” \}


The JSON data format of the BRC-20 deploy transaction includes fields such as protocol name (“p”), operation type (“op”), token identifier (“tick”), maximum supply (“max”), and the maximum number of tokens that can be minted at one time (“lim”). In the above JSON data, “brc-20” indicates the protocol used, “deploy” specifies the operation type, “ordi” is the token identifier, the maximum supply is “21000000”, and the maximum minting quantity at one time is “1000” tokens.

\subsubsection{ The Operation of Mint}

After the token deployment, any user can mint and trade the tokens. Users write the minting function to the Bitcoin chain through commit and reveal transactions to create and distribute a certain number of BRC-20 tokens. The JSON format of a minting transaction is as follows:

\hspace{0.8cm}\{ “p”: “brc-20”,

\hspace{0.8cm} “op”: “mint”, 

\hspace{0.8cm} “tick”: “ordi”, 

\hspace{0.8cm} “amt”: “1000”  \}


	The JSON data format of the BRC-20 mint transaction includes fields such as protocol name (“p”), operation type (“op”), token identifier (“tick”), and the number of tokens to be minted (“amt”). In the above JSON data, “brc-20” indicates the protocol used, “mint” specifies the operation type, “ordi” is the token identifier, and “1000” is the number of tokens to be minted.

\subsubsection{The Operation of Transfer}

Once the total minted tokens equal the total supply, users can only trade or transfer tokens and cannot mint any more. Users write the transfer operation to the Bitcoin blockchain through commit and reveal transactions. Third-party ledgers can recognize transfer operations and obtain transfer addresses and amounts. The JSON format of a transfer transaction is as follows:
 
\hspace{0.8cm}\{ “p”: “brc-20”, 

\hspace{0.8cm} “op”: “transfer”, 

\hspace{0.8cm} “tick”: “ordi”, 

\hspace{0.8cm} “amt”: “500”, 

\hspace{0.8cm} “to”: “bc1q...”, 

\hspace{0.8cm} “from”: “bc1q...” \}


The JSON data format of the BRC-20 transfer transaction must include fields such as protocol name (“p”), operation type (“op”), token identifier (“tick”), transfer amount (“amt”), recipient address (“to”), and sender address (“from”). “brc-20” indicates the protocol used, “transfer” specifies the operation type, “ordi” is the token identifier, “500” is the number of tokens to be transferred, and both recipient and sender addresses must be Taproot-type addresses starting with “bc1q...”.

For BRC-20 token transfers, users first write the transfer operation to the Bitcoin blockchain through commit and reveal transactions. Third-party ledgers can recognize the transfer operations and obtain the transfer addresses and amounts. If the user’s balance in the third-party ledger is greater than the transfer amount, the user can initiate any form of the next transaction to complete the token transfer.
To complete an effective transfer of tokens, the specified amount must not exceed the inscribed $overall$ $balance$. The current $available$ $balance$ is shown as:
\begin{center}
     $Available$ $balance$$=$$Overall$ $balance$$-$$Transferable$ $balance$
\end{center}
 
\noindent Where, $Transferable$ $balance$ is the total tokens held in the user’s wallet. This ensures the executability of token operations within the constraints of the token contract and user token balance. 

The BRC-20 token standard brings new functionalities and applications to the Bitcoin ecosystem, opening a new chapter for Bitcoin in the realm of digital assets. However, the market derived from the BRC-20 token standard relies on centralized off-chain indexers, which conflicts with Bitcoin’s decentralized nature. Effectively resolving this conflict will be key to the long-term development of the BRC-20 ecosystem.

\section{Atomicals Protocol}
\label{sec:Atomicals Protocol}

This section introduces a detailed explanation of the Atomicals protocol, which offers a framework for Non-Fungible Tokens (NFTs) and the ARC-20 token standard\cite{r30}. Similar to the Ordinals protocol, the Atomicals protocol standardizes the mint, transfer, and update operations of digital entities on the Bitcoin blockchain. However, unlike the Ordinals protocol, the Atomicals protocol supports self-certification of digital objects and ownership transfer verification. This means digital objects can automatically verify their authenticity and ownership transfer process on the blockchain, eliminating the need for external third parties or indexers. This enhances the integrity of the digital object history within the blockchain ecosystem, making the Atomicals protocol significantly advantageous in maintaining the reliability and security of digital objects.

\subsection{Digital Object}
The Atomicals protocol provides a framework for minting, transferring, and dynamically updating digital objects (such as non-fungible tokens and fungible tokens) on the Bitcoin blockchain. Compared to the Ordinals protocol, the Atomicals protocol differs in several key aspects.

First, while the Ordinals protocol assigns a unique number to each sat and uses these numbers to identify and track digital objects, the Atomicals protocol assigns a unique Atomicals ID to each digital object. This ID is generated based on the order in which the digital objects are minted, rather than being associated with specific sat.

 Second, the minting process of digital objects, although also achieved through two transactions (commit and reveal) similar to the “inscription” method in the Ordinals protocol, involves different data or file formats embedded in the locking script. Additionally, the Atomicals protocol allows for updating the content of digital objects. This means digital objects can be modified or have new information added later. For example, a digital artwork can have new descriptions, tags, or other metadata added after it is minted. To achieve updates, the Atomicals protocol also uses two transactions, commit and reveal. 
 
\begin{figure}

  \centering
  \setlength{\abovecaptionskip}{0.cm}
  \includegraphics[width=\columnwidth]{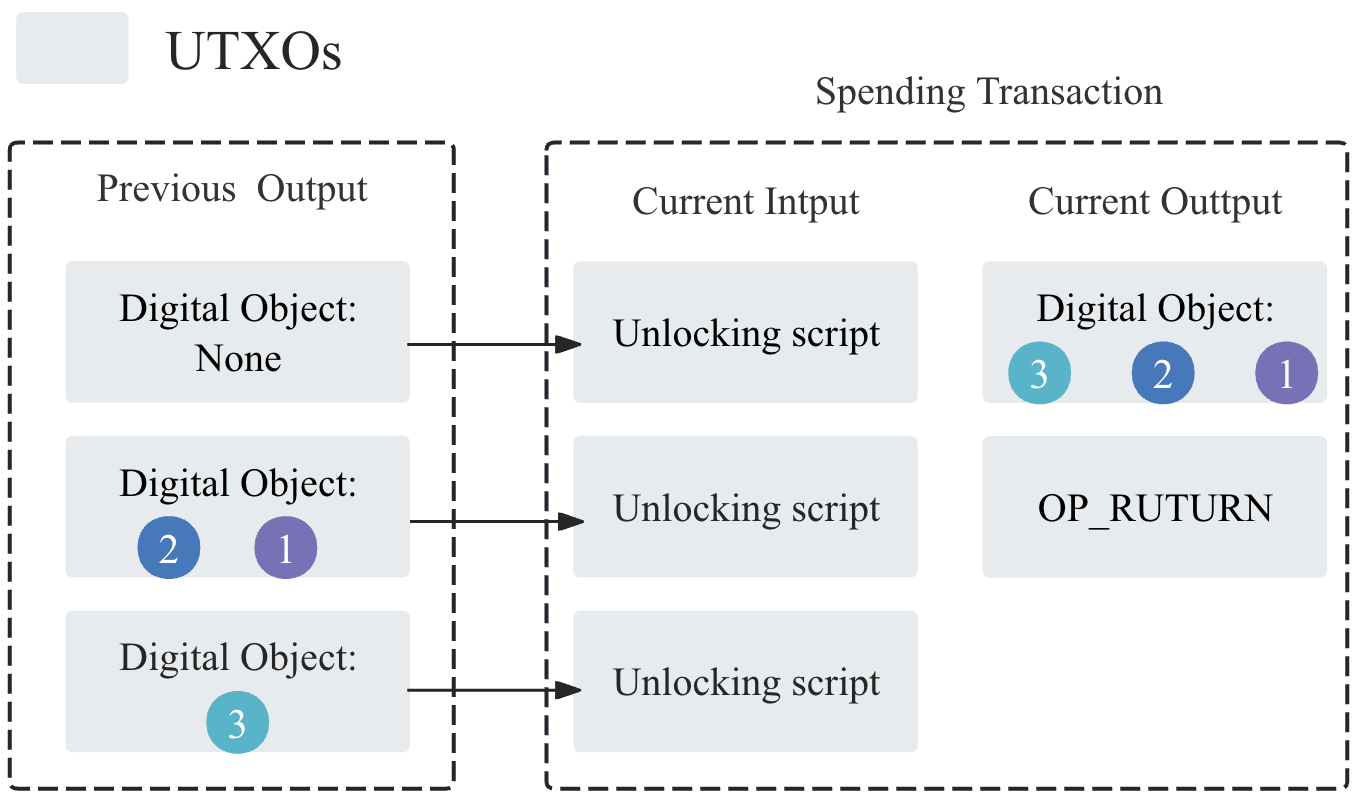}\\
  \caption{The conceptual illustration of the transfer rule in the Atomicals protocol.}\label{FIG. 12}
  \vspace{-0.5cm}  \end{figure}

In terms of transfer rules, the Atomicals protocol is more flexible than the Ordinals protocol. It allows for the use of various address formats, including P2TR, P2SH, and traditional P2PKH addresses. The transfer rules adopt a first-in, first-out (FIFO) method, allocating input digital objects to outputs while ignoring unspendable transaction outputs such as OP\_RETURN. During the transfer process, digital objects from the earliest input are allocated to the first suitable output (excluding OP\_RETURN), followed by the next digital object being allocated to the next output, and so on. If the input digital objects are insufficient to allocate to all outputs, the remaining digital objects are allocated to the first output. As depicted in Fig. \ref{FIG. 12} illustrates the transfer rules of the Atomicals protocol in detail. In the figure, there are three input digital objects (A1, A2, A3) and two outputs (O1, O2), where O2 is an unspendable transaction output OP\_RETURN. According to the FIFO method, A1 is allocated to O1, and since digital objects must ignore unspendable transaction outputs like OP\_RETURN, A2 and A3 are also allocated to O1.

\begin{figure}

  \centering
  \setlength{\abovecaptionskip}{0.cm}
  \includegraphics[width=\columnwidth]{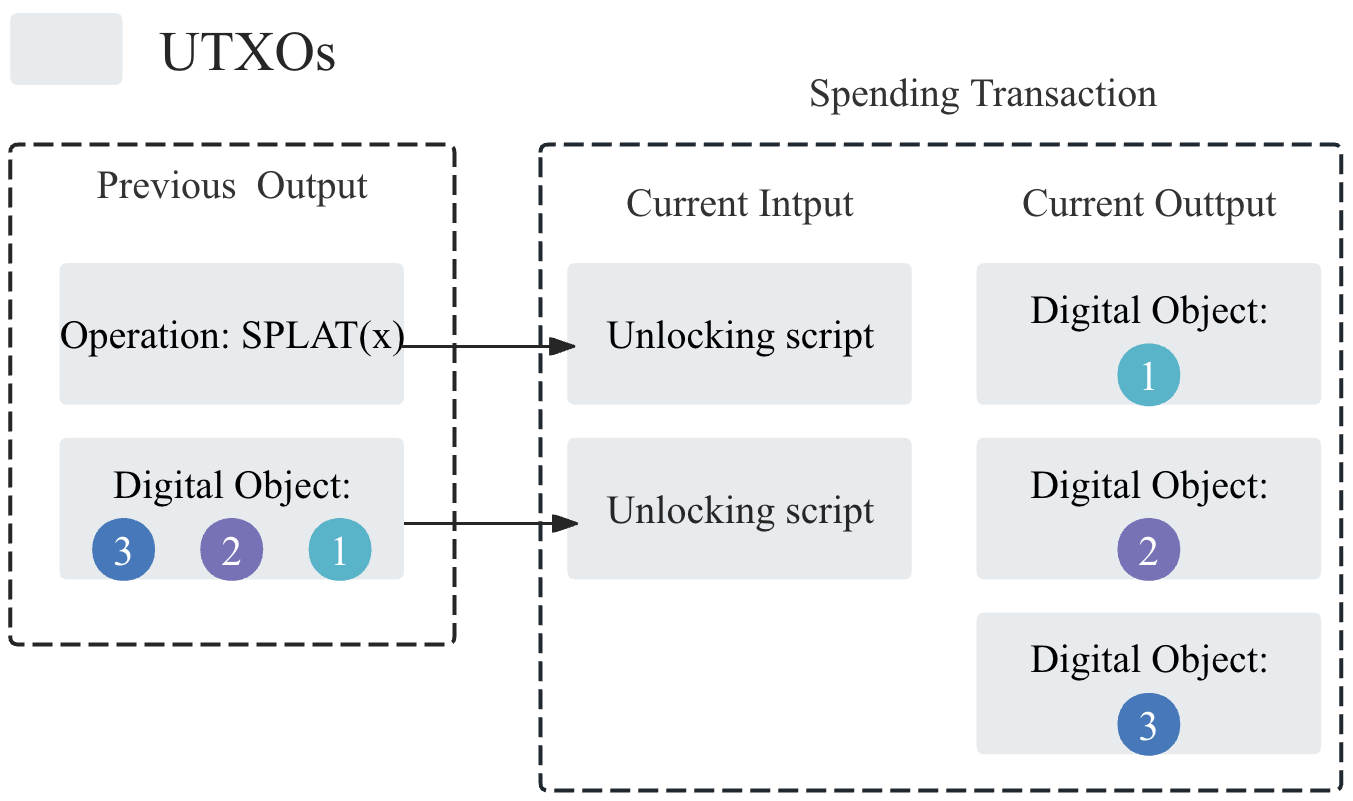}\\
  \caption{The conceptual illustration of the splat operation in the Atomicals protocol.}\label{FIG. 13}
  \vspace{-0.5cm}
\end{figure}

It is important to note that when a transaction includes an unspendable transaction output OP\_RETURN, the digital objects will skip this output and allocate to the next spendable transaction output. This allocation may cause multiple digital objects to merge into a single unspent transaction output. To split these digital objects, the SPLAT(x) operation can be used. This operation distributes the digital objects across multiple outputs, arranged in alphabetical order based on their unique Atomicals IDs. As depicted in Fig. \ref{FIG. 13} details the SPLAT operation in the Atomicals protocol. The purpose of the SPLAT operation is to distribute multiple digital objects from one output to multiple outputs, ensuring each digital object can exist independently and be accurately tracked. In the figure, we can see that the transaction input contains multiple digital objects that need to be distributed across multiple outputs. The SPLAT operation allocates them to different outputs by arranging each digital object’s unique Atomicals ID in alphabetical order. The inputs contain multiple digital objects to be split, each with a unique Atomicals ID; the outputs are multiple target outputs intended to receive the split digital objects.

In summary, although the Atomicals protocol shares some fundamental design similarities with the Ordinals protocol, such as the two-step commit and reveal scheme, it differs significantly in the identification of digital objects, transfer rules, and content updates. The Ordinals protocol assigns a unique number to each satoshi, whereas the Atomicals protocol assigns a unique ID to each digital object, generated based on the order of minting. In terms of transfer rules, the Atomicals protocol allows the use of various address formats and adopts a FIFO method while ignoring unspendable outputs (such as OP\_RETURN). Additionally, the Atomicals protocol supports dynamic updates of digital objects, ensuring the verifiability and immutability of updates through the two-step commit and reveal scheme. These differences make the Atomicals protocol more adaptable and efficient in handling dynamic updates and flexible transfers, making it more suitable for the needs of modern blockchain applications.
\begin{figure*}

  \begin{center}
  \includegraphics[width=\textwidth]{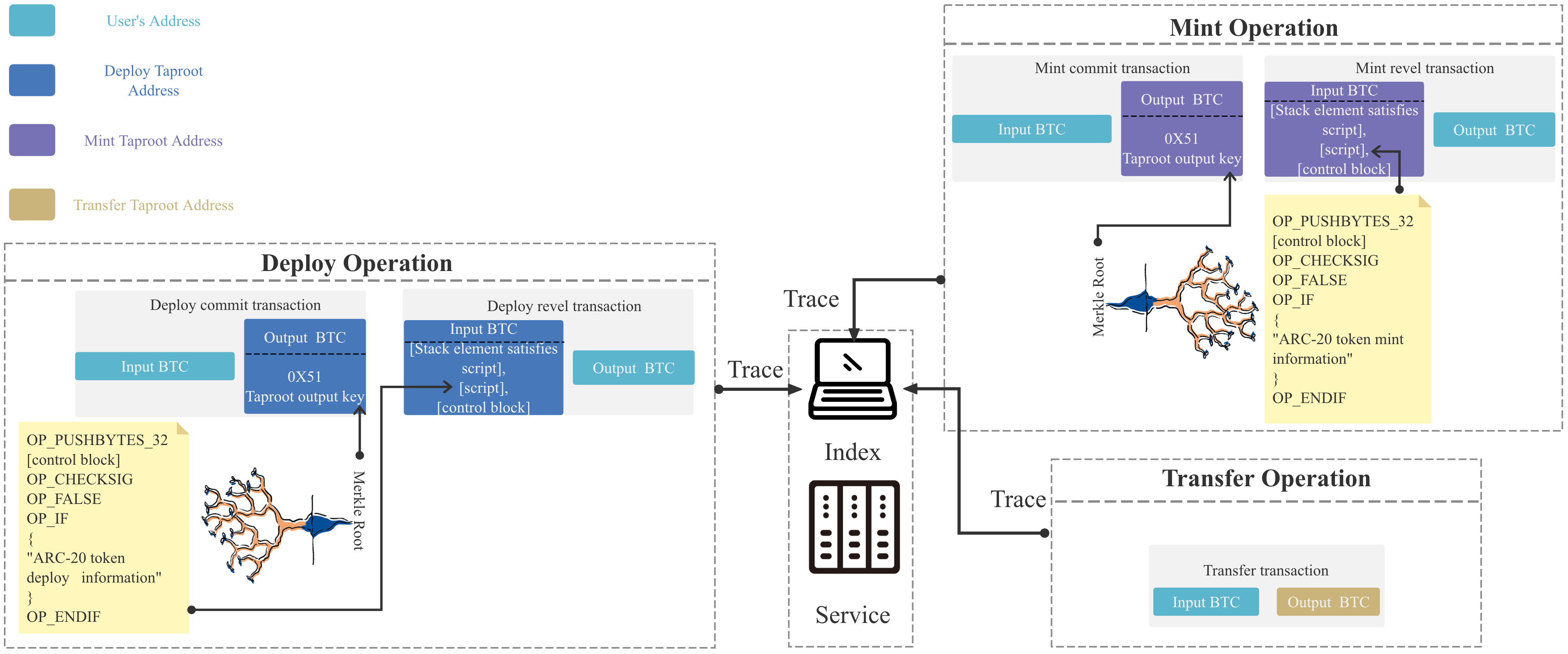}\\
  \caption{The deploy, mint, and transfer processes of ARC-20.}\label{FIG. 14}

  \end{center}
  \vspace{-0.5cm} 
\end{figure*}

\subsection{ARC-20}
The ARC-20 token standard, developed using the Atomicals protocol as the underlying technology, is a fungible token standard. Similar to BRC-20, the ARC-20 token standard enables the creation of fungible tokens on Bitcoin’s base layer. The ARC-20 token standard defines a standardized method for deploy, mint, and transfer tokenized assets on the Bitcoin blockchain. This is achieved by inscribing specific JSON-formatted data onto sats corresponding to the number of tokens. The ARC-20 token standard binds each token to a sat, ensuring that the tokens have a clear value. This design allows ARC-20 tokens to be handled like regular Bitcoin transactions, preserving Bitcoin’s divisibility and composability features. Notably, the ARC-20 token standard does not require additional third-party ledgers to track users’ token balances. This is because the balance of ARC-20 tokens is directly reflected by the number of sats in the UTXO, eliminating the need for external off-chain indexers to record and manage token balances. Each token’s balance and ownership can be verified and transferred through native transactions on the Bitcoin blockchain, avoiding additional centralized dependencies. Additionally, the ARC-20 protocol ensures the scarcity and minting difficulty of tokens through a PoW system called Bitwork Mining. This means that before minting a token, users need to provide proof of work (PoW) for a field named “Bitworkc”. To better illustrate the PoW mechanism in the ARC-20 token standard, we use the example of issuing a fungible token named “Infinity”. This is an infinitely minable PoW-based token whose mining difficulty will continue to increase. Currently, the value of the field “Bitworkc” is “888888888” (its decimal value is 586406201480). Before minting, users need to compute a matching string for “888888888” corresponding to a decimal hash value, with a success probability of 1/586406201480. This is because the hexadecimal string “888888888” converts to a decimal value of 586406201480, meaning that among a total of 586.4 billion possible hash value combinations, only one combination can successfully match this string. Users generate hash values by continuously trying different input data (e.g., different random numbers) until they find a hash value that meets the criteria. Assuming a current market GPU (like the RTX 4090) performs about 3 billion SHA256 calculations per second, a user would need approximately 196 seconds to successfully generate a hash value matching “888888888”. Once the required hash value is generated, users complete the minting process through commit and reveal transactions. As depicted in Fig. \ref{FIG. 14} demonstrates the operational process related to the “Infinity” ARC-20 token, describing its operation of deploy, mint, and transfer on Bitcoin.

The ARC-20 token standard, similar to the BRC-20 standard, defines three token operations: deploy, mint, and transfer. Just like in the BRC-20 token standard, the ARC-20 standard achieves deploy and mint operations through commit and reveal transactions. In these operations, the commit transaction submits specific JSON-formatted data (containing deployment or minting details) to the Bitcoin blockchain via the Taproot output key, but the content is not yet publicly available on the Bitcoin mainnet. The reveal transaction then discloses the deployment or minting details through the [Tapscript] field in the unlocking script. The detailed process is described in Section \ref{sec:Ordinals Protocol and BRC-20 Standard}.

\subsection{Token Standard Comparison}
In this section, we will compare several different Bitcoin token standards, including BRC-20, ARC-20, and the Runes protocol. We will explore their differences in design principles, implementation complexity, and system dependencies to provide a comprehensive understanding of these protocols.

The Runes protocol, proposed by Casey, the creator of the Ordinals protocol, aims to deploy fungible tokens on the Bitcoin blockchain. This protocol addresses the inefficiencies and operational complexities of the BRC-20 token standard and provides a more straightforward and efficient asset issuance and management framework. The issuance method of Runes tokens, called “etching”, binds the balance of each Runes token to a UTXO (unspent transaction output). The OP\_RETURN field of the transaction records data about the balance, token symbol, and other information. OP\_RETURN can be seen as a note for the transaction, with its data marked in JSON format, indicating how many Runes tokens the UTXO represents. To identify these tokens, third-party indexers are required.

The BRC-20 protocol combines the inscription mechanism of the Ordinals protocol, achieving this by inscribing specific JSON-formatted data onto sats. Essentially, this uses BTC as storage space, with transfers entirely reliant on off-chain BRC-20 index ledgers. To ensure the legality of transactions, the transfer of BRC-20 tokens requires commit and reveal transactions to initiate the transfer request, followed by a third transaction to transfer the sats corresponding to the tokens to the target address.

The ARC-20 protocol uses a coloring method for token issuance. In the ARC-20 protocol, the balance of each token is represented by the number of sats in a UTXO (unspent transaction output). This means each token corresponds to a sat, i.e., $1$ token equals $1$ sat. Due to this correspondence, the transfer of ARC-20 tokens can be handled like regular Bitcoin transactions. This design allows ARC-20 tokens to maintain a high degree of decentralization and transparency, avoiding reliance on external indexers.

BRC-20, ARC-20, and the Runes protocol are three different token standards, differing in many aspects. Unlike BRC-20, which uses BTC solely for storage transfer records, the Runes protocol supports a UTXO-based model, leveraging the benefits of UTXOs. This approach is similar to ARC-20 but differs in that UTXOs in the Runes protocol can represent any amounts of tokens, whereas in the ARC-20 protocol, the number of tokens is strictly related to the number of sats in the UTXO. Moreover, ARC-20 relies less on third-party indexers, while the legality of token transfers in BRC-20 and the Runes protocol still needs to be recorded in third-party index ledgers. In terms of the number of transactions required, the BRC-20 protocol involves more deploy, mint, and transfer transactions, while ARC-20 and the Runes protocol involve relatively fewer transactions. TABLE \ref{table I} shows the differences between the three token standards.

\begin{table*}[ht]

\caption{Comparison of Different NF Standards}\label{table I}

\centering
\begin{tabular}{l|cccccc}

Protocol & Based UTXO & Index & The third Ledger & \makecell{Number of deploy \\ transactions} & \makecell{Number of mint \\ transactions} & \makecell{Number of transfer \\ transactions} \\
\hline
  BRC-20 & Not & Yes & Yes & 2 & 2 & 3 \\
  ARC-20 & Yes & Yes & Not & 2 & 2 & 2 \\
  Runes & Yes & Yes & Yes & 2 & 1 & 1 \\

\end{tabular}

\end{table*}

\section{Bitcoin Layer 2 Protocols}
\label{sec:Bitcoin Layer 2 Protocols}

The Ordinals and Automicals protocols write programs directly into the transactions that are stored on the Bitcoin blockchain, so they can be clearly classified as the Layer 1 protocols of Bitcoin. While there is a comprehensive definition of Layer 2 for Ethereum \cite{sguanci2021layer}, Bitcoin lacks a clear definition in this regard. Nevertheless, the Bitcoin community has proposed many other solutions that exploit off-chain mechanisms to extend the programming capability and performance of Bitcoin, even if they have not been formally designated as Layer 2 solutions. Drawing inspiration from Ethereum's Layer 2 definition, we define these off-chain-based protocols in Bitcoin as Layer 2 solutions and categorize them into four distinct types (i.e.,  Rollup, sidechains, client-side verification, and state channel):
\begin{itemize}
    \item {\bf Rollup}: It is a technology that shifts complex transactions or computations off the Bitcoin blockchain for processing and then submits proof of these computations (such as zero-knowledge proofs or optimistic proofs) back to the blockchain for validation. By combining off-chain processing with on-chain validation, Rollup effectively addresses the limitations of block time and block size, enhancing the secure processing capabilities of the blockchain.

    \item {\bf  Sidechains}: It solves Bitcoin’s scalability problem by creating an independent blockchain that runs parallel to the Bitcoin mainchain. These sidechains can generate blocks more quickly and accommodate larger blocks to handle a higher volume of transactions. Sidechains interact with the Bitcoin blockchain through specific bridging mechanisms, such as multi-signature or Isomorphic Binding, allowing for the cross-chain transfer and validation of assets and data.

    \item {\bf Client-side validation}: It is a transaction verification framework that shifts part of the transaction verification logic and process from the Bitcoin blockchain to the client for independent execution, while integrating with the final confirmation steps on the Bitcoin mainchain. Under this framework, the client uses its built-in verification engine to thoroughly review transactions, ensuring they comply with protocol rules. Once verified, the client generates proofs of the results (such as validating digital signatures or checking the sufficiency of transaction balances) and submits them to the Bitcoin blockchain for final confirmation and record-keeping. This method allows users to complete transaction verifications quickly and accurately, even during network congestion, without waiting for block confirmations on the Bitcoin mainchain, thereby addressing Bitcoin’s scalability challenges.

    \item {\bf State channels}: It creates a parallel environment to the mainchain, designed to meet the performance demands of high-frequency, real-time transactions. Represented by the Lightning Network, state channels establish one or more secure off-chain channels that allow participants to complete a large volume of transactions rapidly, only submitting final results to the of Bitcoin blockchain when necessary. This mechanism significantly improves transaction efficiency, particularly in scenarios with frequent, small payments.
    
\end{itemize}

In this section, we present an in-depth analysis of Bitcoin Layer 2 protocols, focusing on BitVM as a representation of Rollup technology. We will examine its core principles and implementation. Additionally, we will present several sidechains protocols, including Babylon, BEVM, RGB++, and STACKS, discussing their underlying concepts and technical frameworks. We will also present client-side validation, with the RGB protocol as a key example, to explain its design and technical characteristics. Finally, we will present the state channel concept, using the Lightning Network as a representative to explain its working principles and technical structure.

\subsection{BitVM}
The BitVM protocol \cite{r9} is based on rollup technology to extend the Bitcoin functionality, which introduces the function of Turing-complete virtual machine to Bitcoin. The virtual machine of BitVM enables the execution of smart contracts within the infrastructure of the Bitcoin blockchain. Due to the constraint imposed by the scripting language and stack-based virtual machine of Bitcoin, the on-chain computation ability of Bitcoin is limited and cannot achieve Turing completeness. The virtual machine of BitVM achieves Turing-complete smart contracts by offloading computationally complex computations to off-chain. As a result, this architecture effectively alleviates the computation burden on the Bitcoin blockchain. Moreover, leveraging the sophisticated methodologies such as Optimistic Rollups for fraud proofs and challenge-response protocols (which originally proposed for Ethereum Layer 2) \cite{r8}, BitVM can maintain security for its off-chain smart contract computations. 

There are two distinct roles in the operational model of BitVM: the prover and the verifier. Together, they are responsible for constructing a program and associating this program to a Taproot address. The computation ability of the constructed program is very broad, encompassing any Turing-complete functions capable of executing complex computational tasks. Both the prover and verifier engage in complex, stateful computations off-chain, without recording the details of these computations on the blockchain. After the prover finishes the execution of the program, the prover needs to make a commintment to the execution process.  The verifier then can send a challenge to the prover for the purpose of verifying whether the prover’s program is correctly executed as commited by the prover. The prover has to send back response to the challenge by revealing the commitment. If the revealed commitment about the prover’s program execution are found to be false, the verifier has the right to sanction the prover. This challenge-response mechanism forces the prover to provide the correct statement. 

\begin{figure}

  \centering
  \setlength{\abovecaptionskip}{0.cm}
  \includegraphics[width=\columnwidth]{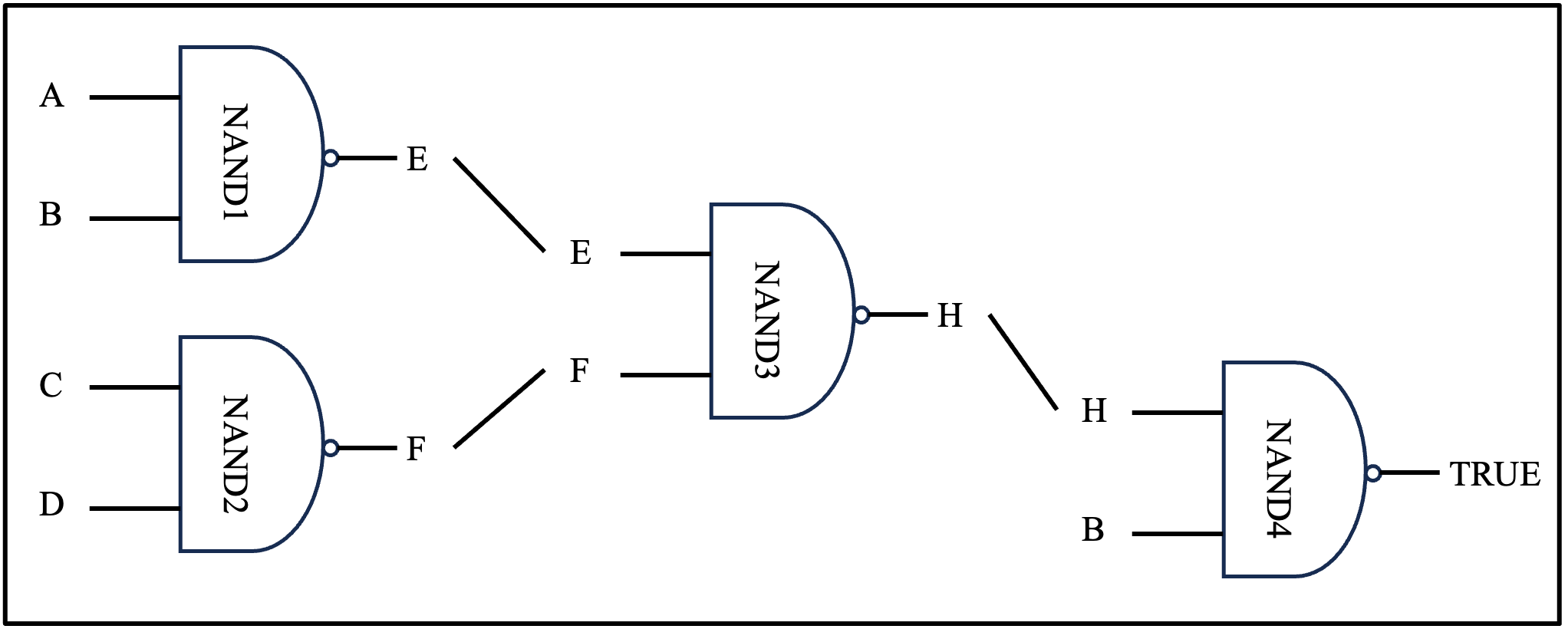}\\
  \caption{An example of binary circuit to represent the program execution process.}\label{FIG. 15}

  \vspace{-0.3cm} \end{figure}

Under the BitVM protocol, the execution of the program is expressed as a binary circuit. In Fig. \ref{FIG. 15}, we give a simple binary circuit as the example A binary circuit is composed of logic gates that perform AND, OR, and NOT operations. Notably, the logic gate of the NAND (NOT AND) operation can be used as a universal logic gate capable of constructing the logic gates of all other operations \cite{r35}. Therefore, we can construct binary circuits using only NAND gates. Meanwhile, within Bitcoin’s scripting language, a combination of the opcodes OP\_BOOLAND and OP\_NOT can facilitate the creation of a new opcode, “OP\_NAND”, which can perform the NAND operation. For example, the binary circuit given in Fig. \ref{FIG. 15} consists of four NAND gates. Each NAND gate contains two inputs and an output, and the state of each input/output is represented as a bit variable that takes a value of 0 or 1.

Now, to make a commitment to the execution process of the program, the prover can make commitment to the binary circuit, called Binary Circuit Commitment, which is sued ensure the correctness and completeness of the program execution process. A Binary Circuit Commitment consists of several Logic Gate Commitment, and each Logic Gate Commitment corresponds to each logic gate in the binary circuit. For example, the Binary Circuit Commitment of the binary circuit in Fig. \ref{FIG. 15} consists of four Logic Gate Commitments. Moreover, each Logic Gate Commitment contains three Bit Value Commitments, corresponding to the state bit variables of the two inputs and one output of the logic gate. These commitments are public to the verifier, who can use them to verify the correctness of the program execution. In the following, we will explain what Bit Value Commitment, Logic Gate Commitment and Binary Circuit Commitment are, how to use Bitcoin scripts to implement them, and how to use them to implement the challenge-response mechanism of BitVM.

\subsubsection{ Bit Value Commitment}

Bit value commitments are fundamental constructs in the BitVM protocol. It enables a prover to assign a bit variables of “0” or “1” to the two inputs and one output of the logic gate. The following Bitcoin script is about to create a UTXO that embodies a Bit Value Commitment.

\begin{algorithm}

OP\_IF\hspace{3cm} /*An opcode used to execute con-

\hspace{4.2cm}ditional branches in Bitcoin 

\hspace{4.2cm}scripts. */

\hspace{1cm} OP\_HASH160 \>	\hspace{0.4cm} /*An opcode used to compute 

    \hspace{4.2cm}the RIPEMD-160 hash of the 
    
    \hspace{4.2cm}top stack item. */

\hspace{1cm}\textless HashA1\textgreater \hspace{1.4cm}/* A placeholder for the hash 

\hspace{4.2cm}value, representing the HashA1 

\hspace{4.2cm}value.*/

\hspace{1cm}OP\_EQUALVERIFY	 /* An opcode used to compare

\hspace{4.2cm}the top two stack items for

\hspace{4.2cm}equality and verifies the 

\hspace{4.2cm}result as true. */

\hspace{1cm}\textless 1\textgreater 
  
OP\_ELSE \hspace{2.6cm}/* Same as before. */

\hspace{1cm}OP\_HASH160

\hspace{1cm}\textless HashA0\textgreater \hspace{1.5cm}/*A placeholder for the hash 

\hspace{4.2cm}value, representing the HashA0 

\hspace{4.2cm}value. */

\hspace{1cm}OP\_EQUALVERIF

\hspace{1cm}\textless 0\textgreater

 OP\_ENDIF \hspace{2.4cm}/*The end marker of the condi-
 
 \hspace{4.2cm}tional statement. */
 
\end{algorithm} 

First, the prover selects two preimages, representing the values “0” or “1” of the state bit variables for the logic gate's input/output. These are denoted as \textless PreimageA0\textgreater and \textless PreimageA1\textgreater. The corresponding hash values, \textless HashA0\textgreater and \textless HashA1\textgreater, are computed using the OP\_HASH160 opcode in the Bitcoin script. Then, the prover generates a Bit Value Commitment by creating a locking transaction for a UTXO containing a script that includes \textless HashA0\textgreater and \textless HashA1\textgreater. This script can be abstracted as “\textless HashA0\textgreater \textless HashA1\textgreater OP\_BITCOMMITMENT”.

After that, the prover reveals the preimage corresponding to either \textless HashA1\textgreater or \textless HashA0\textgreater to the verifier by fulfilling the unlocking conditions of the UTXO containing the Bit Value Commitment. The verifier then verifies the correctness of the prover's Bit Value Commitment by using the script-locked \textless HashA0\textgreater or \textless HashA1\textgreater along with the revealed preimage to check the integrity of the commitment. The entire verification process can be performed by cascading the preimages of \textless HashA0\textgreater or \textless HashA1\textgreater with the Bitcoin script used to create the Bit Value Commitment.

\subsubsection{Logic Gate Commitment}

In the BitVM protocol, a Logic Gate Commitment consists of three Bit Value Commitments corresponding to the two inputs and one output of a logic gate. These Logic Gate Commitments are further interconnected to form a Binary Circuit Commitment. This method transforms the program's execution into a structured format that is both compatible with and executable within Bitcoin Script.

Assume the inputs of the logic gate are A and B, and the output is C. The prover generates a Logic Gate Commitment using the following Bitcoin script:

\begin{algorithm}

\textless HashC0\textgreater	\textless HashC1\textgreater OP\_BITCOMMITMENT	

OP\_TOALTSTACK	\hspace{0.8cm}/* An opcode used to move top 

\hspace{3.7cm}stack element from the main stack 

\hspace{3.7cm}to the altstack. */

\textless HashB0\textgreater \textless HashB1\textgreater OP\_BITCOMMITMENT	 

OP\_TOALTSTACK	\hspace{0.8cm}/* Same as before. */

\textless HashA0\textgreater \textless HashA1\textgreater OP\_BITCOMMITMENT 

OP\_FROMALTSTACK \>	/* An opcode used to move the top

\hspace{3.7cm}stack element from the alternative

\hspace{3.7cm}stack back to the main stack. */

OP\_NAND	\hspace{2cm}/* An opcode used to perform a  

\hspace{3.7cm}logical NAND operation on the top 

\hspace{3.7cm}two elements of the stack. */

OP\_FROMALTSTACK \>		 /* Same as before. */

OP\_EQUALVERIFY\> \hspace{0.4cm}	/* An opcode used to compare the 

\hspace{3.7cm}top two elements of stack, and if 

\hspace{3.7cm}they are equal, the script continues 

\hspace{3.7cm}execution. If not, the script termina-

\hspace{3.7cm}tes immediately. */
\end{algorithm}

In this script, three Bit Value Commitments are combined with an OP\_NAND script to generate a Logic Gate Commitment. This script can be summarized as “\textless HashC0\textgreater \textless HashC1\textgreater \textless HashB0\textgreater \textless HashB1\textgreater \textless HashA\\0\textgreater 
\textless HashA1\textgreater OP\_GATECOMMITMENT”. Similar to Bit Value Commitments, the prover must reveal the preimages for \textless HashA0\textgreater/\textless HashA1\textgreater, \textless HashB0\textgreater/\textless HashB1\textgreater, and \textless HashC0\textgreater/\textless HashC1\textgreater to the verifier. Finally, the verifier performs an off-chain verification of the Logic Gate Commitment based on the revealed preimages. If the verification process results in a true value at the top of the stack, the logic statement $A$ $NAND$ $B$ $==$ $C$ is confirmed.

\subsubsection{Binary Circuit Commitment}

In the previous section, we defined Logic Gate Commitments and provided an example of generating them using NAND gates. Since any binary circuit can be represented by a combination of NAND gates, the binary circuit of the program execution shown in Fig. \ref{FIG. 15} can be represented by four NAND gates. Consequently, we can use four Logic Gate Commitments to represent the Binary Circuit Commitment.

\begin{figure*}

  \begin{center}
  \includegraphics[width=\textwidth]{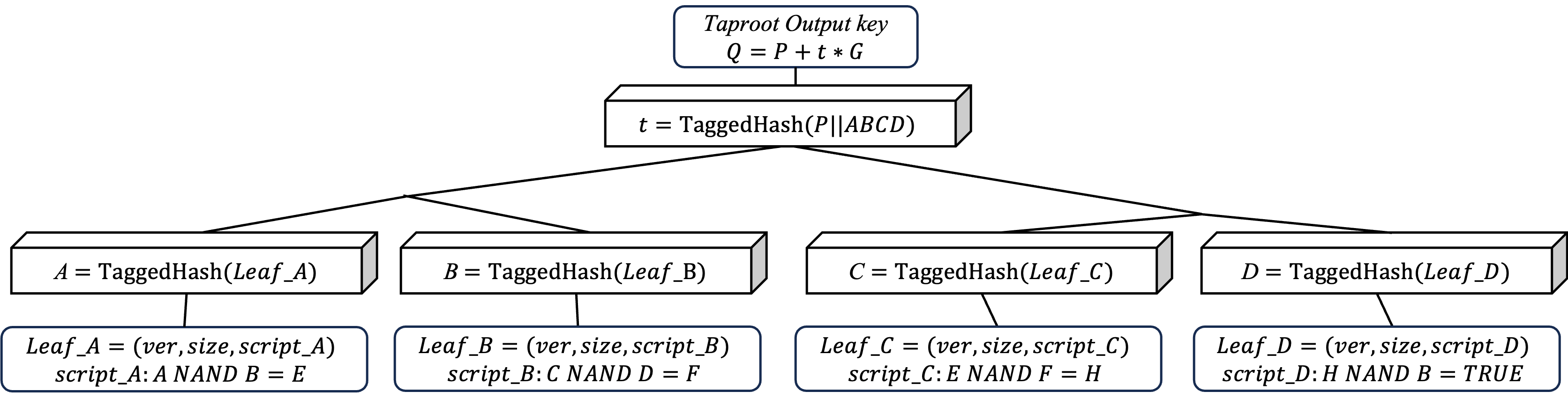}\\
  \caption{The illustration of the Taproot commitment that represents the Binary Circuit Commitment.}\label{FIG. 16}
  \end{center}
  \vspace{-0.5cm} 
\end{figure*}

In practical applications, the corresponding binary circuit typically consists of many logic gates, potentially numbering in the thousands. To optimize on-chain resource utilization and minimize associated costs, the prover can place the scripts for these Logic Gate Commitments in the leaf nodes of a Taproot tree structure. Then, by performing a series of calculations on the leaf nodes, we can obtain the Taproot output key $Q$. This method combines the scripts of the Logic Gate Commitments into the Taproot Commitment and stores it within the Taproot output key. This approach leverages the efficiency of the Taproot structure to manage complex operations while maintaining the integrity and cost-effectiveness of on-chain transactions. Fig. \ref{FIG. 16} provides an example of a Taproot Commitment representing a Binary Circuit Commitment.

\subsubsection{Pre-sign Off-chain Transactions}

In Bitcoin, normal transactions need to be broadcast to the network for confirmation immediately after they are created and signed by the user. However, pre-signed transactions are handled differently from normal transactions. A pre-signed transaction is one that has been signed in advance but not immediately broadcast. It will be included in a new block only when certain conditions (such as time locks) are met.

In the BitVM protocol, the prover and verifier generate these pre-signed transactions using joint signatures to ensure that the script content in the transaction inputs and outputs conforms to the specific format requirements of each step in the protocol. Specifically, in BitVM, a series of challenge or response transactions required when the verifier questions the prover's claims are jointly signed off-chain by both the prover and verifier. These transactions are linked together through the challenge and response mechanism. Joint signatures ensure that each transaction is correctly structured, maintaining the operational integrity of the protocol.

\subsubsection{Challenges and Responses}

\begin{figure*}

  \begin{center}
  \includegraphics[width=\textwidth]{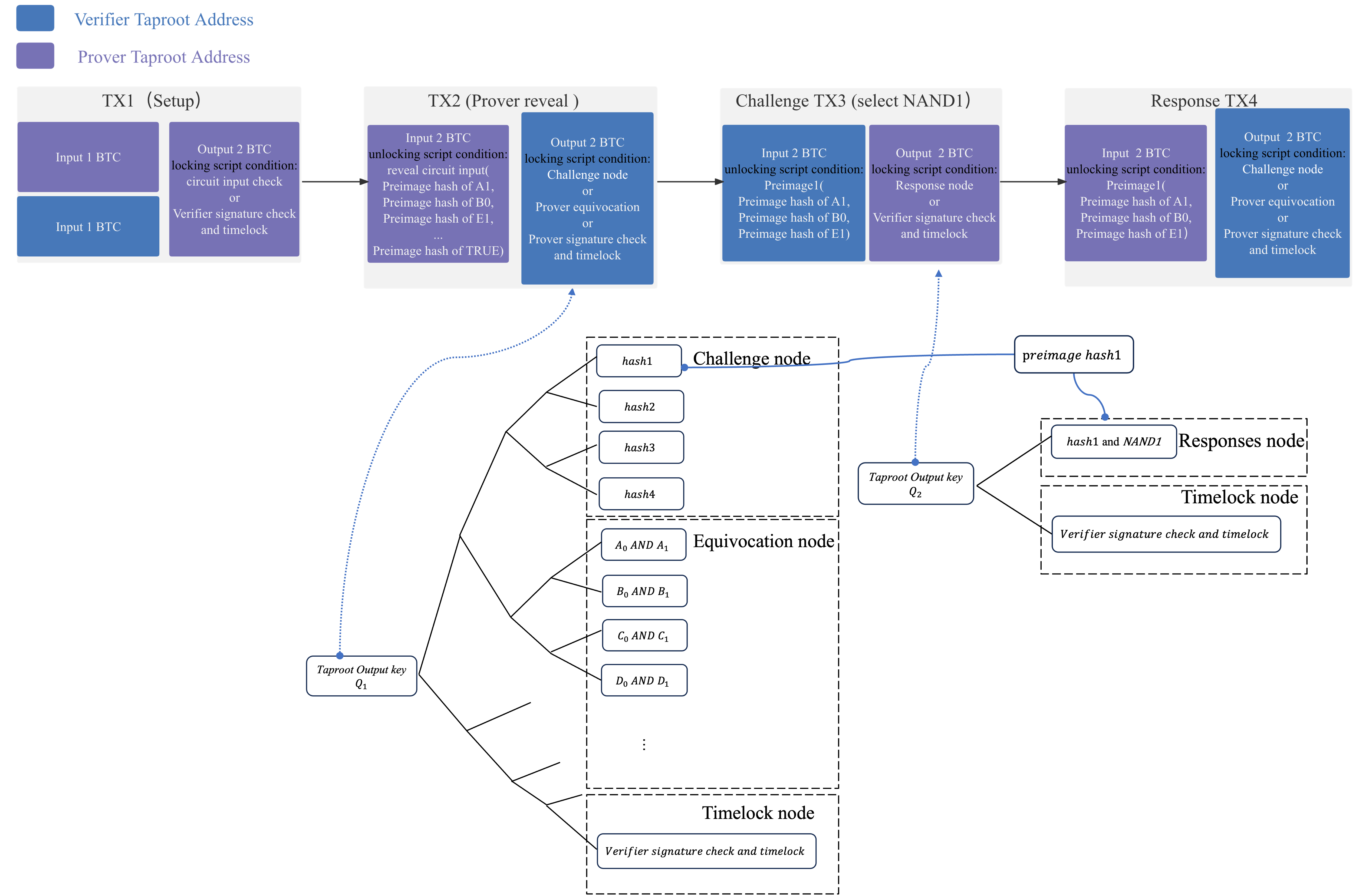}\\
  \caption{The illustration of the challenge and response process in BitVM.}\label{FIG. 17}
  \end{center}
  \vspace{-0.5cm} 
\end{figure*}

In the BitVM protocol, if the verifier has no doubts about the binary circuit execution committed by the prover, they can execute any agreed-upon operation using a 2-of-2 multisig mechanism. This includes contract settlement, fund transfers, and state updates. For instance, the prover and verifier can jointly sign a transaction to close a payment channel and submit it to the Bitcoin mainnet for confirmation. However, if the verifier questions the prover's binary circuit commitment, the challenge and response mechanism is triggered. In this process, if the verifier discovers inconsistencies in the prover's binary circuit execution claims, the verifier is entitled to seize the prover's collateral. The entire process involves several steps, including jointly initiating the binary circuit setup transaction, the prover revealing preimages, the verifier initiating a challenge, and the prover responding to the challenge, as illustrated in Fig. \ref{FIG. 17}. First, both parties jointly initiate the binary circuit setup transaction (TX1: setup). This transaction locks both parties' funds on the mainnet and creates the collateral, laying the foundation for the challenge and response mechanism. This setup specifies that when the verifier challenges the prover, the prover must reveal all preimages involved in the binary circuit to validate the correctness of their claims. Next, the prover sends a reveal transaction (TX2: prover reveal) that reveals all the preimages associated with the binary circuit commitment, including all the bit commitments of inputs and outputs involved in each logic gate.

The verifier then sends a challenge transaction (TX3: Challenge), selecting a specific NAND logic gate to challenge by requesting the preimages of the inputs and outputs of that gate. Finally, the prover responds with a transaction (TX4: Response) that provides the preimages of the challenged NAND gate's inputs and outputs to prove the correctness of their computation. The verifier validates the commitment based on the provided preimages. If the verification succeeds, the process may continue with further challenges and responses; otherwise, the verifier may confiscate the prover's collateral.

This challenge and response cycle repeats until all logic gates are verified or the verifier ceases to challenge the prover. In the following, as depicted in Fig. \ref{FIG. 17}, we will detail each step of the challenge-response mechanism and the associated transactions. For simplicity, we omit the typical miner fees associated with each transaction.

Initially, the verifier and prover jointly initiate the Setup transaction. In this transaction, both parties agree to deposit a certain amount of BTC as collateral (in the example shown in Fig. \ref{FIG. 17}, each party deposits 1 BTC), and output a Taproot-type locking script. During the challenge and response mechanism, if the prover is found to act maliciously, the verifier can claim all the collateral as a penalty. Conversely, if the verifier does not detect any malicious behavior by the prover during the process, the prover can claim all the funds as a reward. This mechanism ensures that the prover completes the challenge and response process on time, or else they face financial loss.

The Taproot locking script output includes two unlocking conditions. The first unlocking condition requires the prover to provide the preimages of all inputs and outputs of the logic gates in the binary circuit. As shown in Fig. \ref{FIG. 15}, the inputs and outputs of the logic gates in the binary circuit include A, B, C, D, E, F, H, and TRUE. In other words, some leaf nodes in the MAST are the bit commitments for all inputs and outputs in Fig. \ref{FIG. 16}. The second unlocking condition is a time lock, meaning that after a certain period, if the prover fails to provide the necessary preimages to spend the UTXO (i.e., does not demonstrate any inaccuracies in the verifier's committed computation), the verifier can unlock the UTXO using only their signature to claim the BTC collateralized in the transaction.

Next, the prover generates a reveal transaction to unlock the locking script created by the setup transaction. To avoid forfeiting the collateralized BTC, the prover must provide the preimages of all logic gates in the binary circuit as part of the unlocking script in the reveal transaction. This satisfies the unlocking condition of the UTXO created by the setup transaction and generates a new UTXO. The locking script for this new UTXO is of Taproot type, with the MAST's leaf nodes consisting of three parts. The first part of the MAST's leaf nodes includes the commitments of all logic gates in the binary circuit (as in Fig. \ref{FIG. 15}), placed in different leaf nodes of the MAST. This allows the verifier to select and unlock the preimages of the inputs and outputs of a logic gate provided by the prover.The second part is activated if the verifier detects a contradiction in the prover's revealed content. For example, the prover reveals the output value on D from the challenged NAND1 gate as D equals 1. However, the verifier finds that the value is D equals 0 upon validating the prover's pre-image, constituting a contradiction. In such cases, the verifier can unlock the script by obtaining two different preimages of D and claim all funds. The third part is a time lock. If the verifier does not further challenge the prover's revelation, the prover can transfer the funds with their signature after a certain period.

If the verifier continues to doubt the prover, they can challenge the inputs and outputs of any logic gate in the binary circuit. The verifier assembles the preimages of the inputs and outputs of the selected gate into an unlocking script. This script satisfies the unlocking condition of the UTXO created by the second transaction. Using this unlocking script, the verifier generates a challenge transaction to challenge the commitment of the selected logic gate.

The locking script of the response transaction consists of a MAST. The leaf nodes of the MAST are made up of two parts. The first unlocking condition of the script requires the prover to use the preimages corresponding to the logic gate selected by the verifier to unlock it. The second part of the locking script is a time lock. If the prover does not initiate a response transaction in time, the verifier can transfer all the funds using their signature after a certain period.

Finally, the prover issues a response transaction to address the challenge transaction. In this response transaction, the prover uses the preimages corresponding to the logic gate selected by the verifier as the unlocking script, satisfying the unlocking condition of the UTXO created by the third transaction. The composition of the new locking script's MAST leaf nodes is identical to that in the second transaction. The first part allows the verifier to select and challenge the revealed preimages of the logic gate, the second part allows the verifier to transfer funds if a contradiction is found in the prover's response, and the third part allows the prover to transfer all funds with their signature after a certain period.

If the verifier continues to doubt the prover, both parties can repeat the challenge and response transactions multiple times until the verifier is satisfied or all logic gate challenges are completed, after which the verifier accepts the prover's commitment as correct. Throughout this process, the verifier will keep challenging the prover, and the prover will respond with evidence to support the correctness of their binary circuit commitment. However, if a contradiction is found during the challenge and response process, indicating potential malicious behavior by the prover, the verifier can immediately generate a new transaction to transfer the funds as a penalty to the prover.

\subsection{Babylon}
In January 2022, Fisher and Professor David Tse from Stanford University co-founded the Babylon project. This project aims to explore the sharing of Bitcoin's security\cite{r39}\cite{r38}. The vision of the project is to leverage Bitcoin’s Proof-of-Work (PoW) security to enhance the security of other blockchains based on Proof-of-Stake (PoS) consensus mechanisms. In a PoS chain\cite{r13}, participants secure the network by staking capital. To validate blocks, participants must lock a certain amount of tokens as collateral. This staking mechanism incentivizes participants to become validators, honestly validate blocks, and maintain blockchain security. When validators stake more funds, they have a stronger economic incentive to protect and support the network. As the staked amount increases, attacks become more expensive and difficult because attackers would need to acquire or rent more tokens to control the network. Thus, the staking mechanism in PoS systems strengthens the overall security of the network.In PoS-based blockchains like Ethereum, Cosmos, and BNB Chain, security is typically ensured by staking their native tokens (e.g., ETH, ATOM, BNB). However, relying solely on native tokens for security limits the PoS chain's security to the total market value of its native tokens. If the total market value of the native tokens is low, attackers can more easily accumulate enough tokens to launch an attack, threatening network security. This limitation makes the security of PoS chains heavily dependent on the market value of their native tokens, and market volatility can reduce security. To address this issue, a method known as “cross-chain staking” was recently proposed\cite{r40}. In this approach, foreign assets remain stored on their original chain but are locked in staking contracts and delegated to specific validators on the target chain. The staked assets are only slashed when the validator commits a slashing offense.

The Babylon project, on the other hand, introduces a breakthrough Bitcoin staking protocol. This protocol allows Bitcoin holders to stake directly on PoS blockchains without intermediaries, bridges, third-party custody, or oracles. The Babylon project has developed two mechanisms for cross-chain staking: the “Bitcoin Timestamp Protocol” and the “Bitcoin Staking Protocol,” which are introduced as follows.

{\bfseries Bitcoin Timestamp Protocol\cite{r37}}:
This protocol generates concise and verifiable timestamps for PoS chain block data on the Bitcoin blockchain. The Bitcoin Timestamp Protocol effectively addresses the “Long-range Attack” security challenge faced by PoS chains\cite{r35}. In a blockchain, a Long-range Attack is a theoretical threat that can undermine the PoS consensus. Similar to a 51\% attack on PoW blockchains \cite{shi2022pooling}, a Long-range Attack allows attackers to create a false chain by starting from the genesis block with a higher token balance than the forging balance and deceiving nodes by creating their own chain branch. Some PoS chains, including Ethereum, use social consensus mechanisms to counteract this attack. In this mechanism, stakeholders periodically make off-chain decisions about the correct block at the latest height and ignore any other potential forks. However, this approach relies on the subjective opinions of stakeholders, compromising decentralization. Moreover, social consensus takes time, so most PoS chains impose long unbonding periods (usually several weeks) to ensure validators cannot easily exit before consensus is reached. The Babylon project introduces the Bitcoin Timestamp Protocol, which timestamps PoS chain block data on the Bitcoin blockchain. In the event of a fork caused by a Long-range Attack, the blocks generated by the attack would have later timestamps on the Bitcoin blockchain compared to normal blocks. Nodes can easily detect these forged blocks, rendering the attack ineffective.

\begin{figure}

  \centering
  \setlength{\abovecaptionskip}{0.cm}
  \includegraphics[width=\columnwidth]{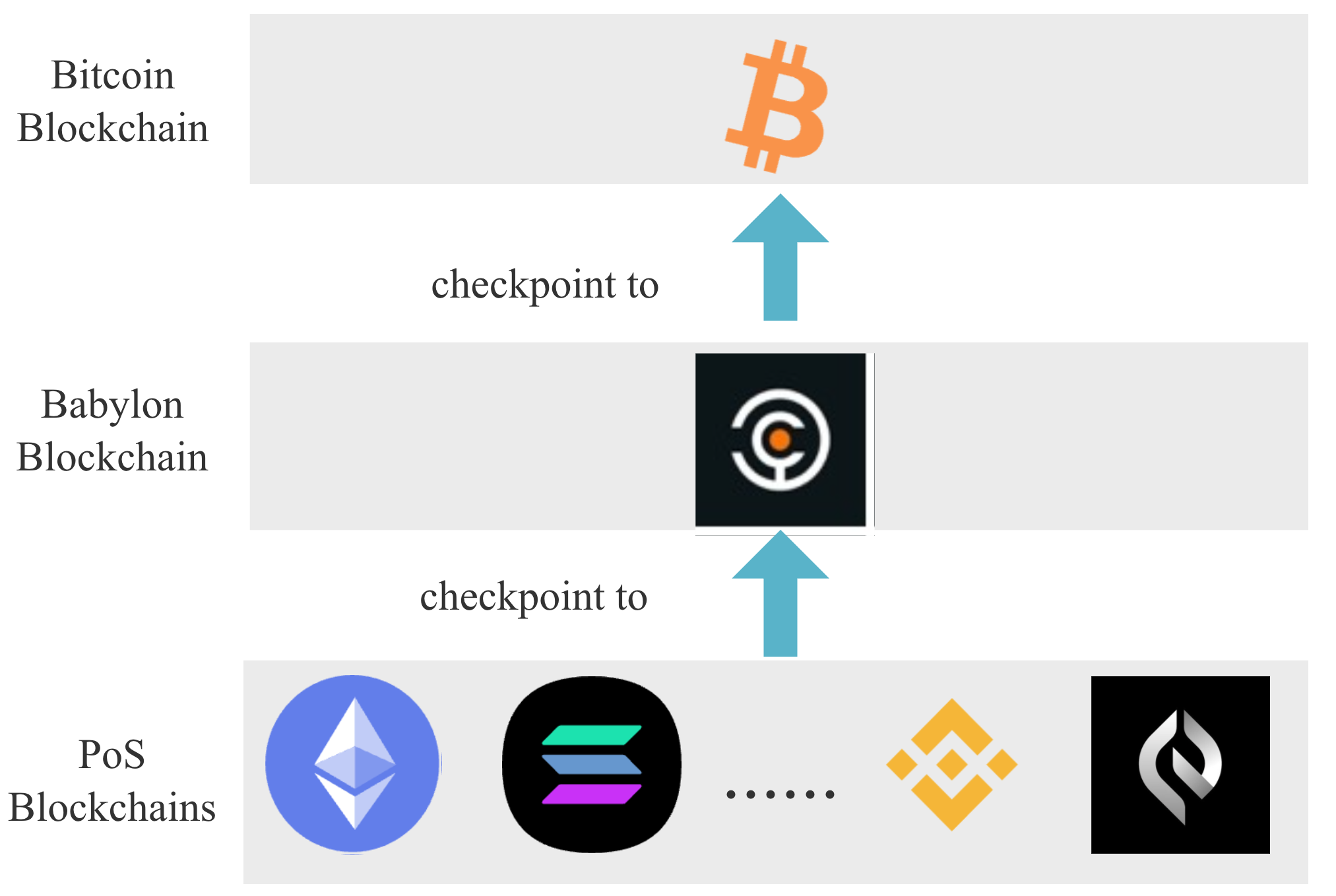}\\
  \caption{The architecture of the Babylon project.}\label{FIG. 18}

  \vspace{-0.5cm} \end{figure}

{\bfseries Bitcoin Staking Protocol\cite{r36}}:
This protocol allows Bitcoin assets to provide economic security to any decentralized system through trustless (self-custody) staking. The Babylon project uses a remote staking method. It locks the staked Bitcoin in a contract on the Bitcoin chain and slashes the staked Bitcoin on the Bitcoin chain when the staker violates the PoS chain protocol. However, Bitcoin’s scripting language has limited expressiveness and cannot support smart contracts. The Babylon project overcomes this challenge by combining advanced cryptographic techniques (extractable one-time signature), consensus protocol innovations, and optimized use of Bitcoin's scripting language. This enables the protocol to slash the staked BTC on the Bitcoin blockchain in case of a violation. The working principle of the Bitcoin Staking Protocol is as follows:
\begin{itemize}

\item Stake Bitcoin: The staker initiates a staking transaction on the Bitcoin blockchain. This transaction creates a UTXO with two spending conditions: 1) a time lock, allowing the staker to retrieve the staked funds after a certain period using their key, and 2) a condition allowing the validator to destroy this UTXO using a special “extractable one-time signature” (EOTS) in the event of a violation by the staker. If delegated, this EOTS belongs to the validator to whom the staker has delegated.

\item Validation on the PoS Chain: Once the staking transaction is confirmed on the Bitcoin chain, the staker (or the validator to whom the staker has delegated) can start validating the PoS chain and sign votes for valid blocks using the EOTS key.
\end{itemize}
To implement these mechanisms, the Babylon project is architected as shown in the Fig. \ref{FIG. 18}, consisting of three parts:

\begin{itemize}
\item [1.] Bitcoin Blockchain: In the Babylon project, the Bitcoin blockchain provides timestamping services to PoS chains, offering reliable timestamps for other blockchains. Additionally, Bitcoin serves as staked capital, leveraging its high market value to provide robust security for PoS chain networks.

\item [2.] Babylon Blockchain: This serves as an intermediary layer and the control plane. Babylon Chain is built on the Cosmos SDK and links Bitcoin and other PoS blockchains through the standard Inter-Blockchain Communication (IBC) protocol. Its primary function is to efficiently aggregate timestamps generated from PoS chain block data. This approach is necessary because the limited and expensive block space on the Bitcoin blockchain makes it unsustainable to individually and continuously generate timestamps for each PoS chain block on the Bitcoin blockchain.

\item [3.] PoS Blockchains: In the Babylon project, other PoS blockchains (such as Ethereum) act as security consumers. They use the Bitcoin security provided by Babylon Chain to enhance their own security. These PoS blockchains rely on the Proof-of-Stake mechanism to reach consensus and use staked capital as an economic incentive to protect and maintain network security.
\end{itemize}

In the Babylon project, participants can choose their roles based on their preferences, including Bitcoin holders, validator nodes, and delegators. Bitcoin holders can lock and stake their Bitcoin to provide economic security to the system, earning returns or rewards in the process. Validator nodes are responsible for verifying the legitimacy of transactions and blocks on the PoS chain. They need the technical ability and resources to perform these validation tasks. Delegators, who may lack the capacity or knowledge to run validator nodes, can delegate their Bitcoin to validators. By doing so, they transfer their voting power to the validator, who will then validate transactions and blocks on the PoS chain on their behalf.

The Bitcoin staking protocol in the Babylon project offers three key security properties. First, it provides fully slashable PoS security, meaning that if the integrity of the blockchain is compromised, one-third of the staked Bitcoin will be slashed. As long as two-thirds of the staked Bitcoin adheres to the PoS protocol, the PoS chain can maintain good liveness. Second, the security of stakers is assured—each Bitcoin staker can retrieve or unbond their staked Bitcoin at maturity, as long as they honestly follow the PoS protocol. Finally, the liquidity of stakers is guaranteed, allowing for rapid unstaking without requiring community consensus. A key design principle of the Bitcoin staking protocol is to make it as close as possible to the native staking mechanism of PoS chains, including PoS token holders, validator nodes, and delegators. For BTC staking, Bitcoin holders can become validators or delegate their Bitcoin to validators. This design allows stakers to participate in staking in a manner similar to PoS chains, thereby providing economic security.

\subsection{BEVM}
We classify the BEVM as a sidechain Bitcoin Layer 2 scaling solution. Due to the Taproot upgrade introducing the Schnorr signature algorithm, its linearity made multisignature schemes for Bitcoin feasible, offering a multisignature solution for Bitcoin's Layer 2 protocols without increasing script size. The Bitcoin-Ethereum Virtual Machine (BEVM), a Bitcoin EVM, will be launched in 2024\cite{r21}. It is an EVM-compatible, fully decentralized Bitcoin Layer 2 protocol that uses BTC as gas. BEVM allows users to transfer their Bitcoin assets from the Bitcoin blockchain to the BEVM chain in a fully decentralized manner. Additionally, BEVM assets and data can be cross-chained back to the Bitcoin blockchain in a decentralized way. Since BEVM is compatible with the Ethereum Virtual Machine (EVM)\cite{r32}, Ethereum ecosystem decentralized applications (DApps) can run on Bitcoin's Layer 2, increasing BTC consumption scenarios and enabling decentralized Bitcoin financial services.
\begin{figure}

  \centering
  \setlength{\abovecaptionskip}{0.cm}
  \includegraphics[width=\columnwidth]{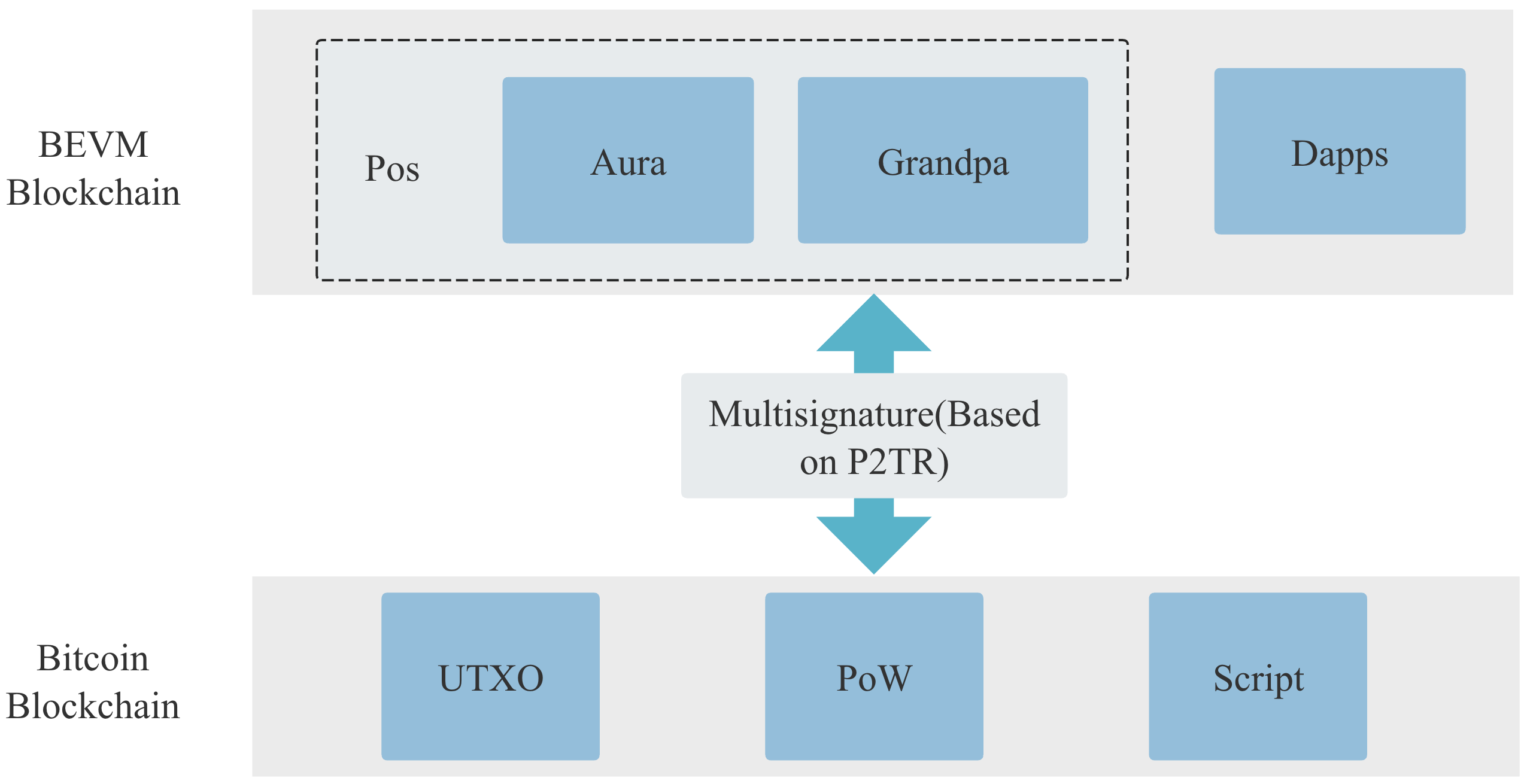}\\
  \caption{The architecture of BEVM.}\label{FIG. 19}
  \vspace{-0.5cm} \end{figure}

As shown in Fig. \ref{FIG. 19}, the BEVM chain uses the Proof of Stake (PoS) mechanism to generate blocks and employs the Aura consensus algorithm to elect block-producing nodes. Here’s how it works: First, a group of PoS consensus nodes is elected based on the amount of stake they hold. These nodes, selected according to their stake, take turns generating blocks in a round-robin order. During block generation, the BEVM chain uses the Grandpa algorithm \cite{r15} to achieve BFT consensus. Grandpa finalizes consensus through validator votes, ensuring that the generated block is widely accepted by the network and becomes an irreversible part of the chain.

To transfer a user's Bitcoin assets from the Bitcoin blockchain to the BEVM chain, this solution synchronizes Bitcoin light node data on the BEVM chain. Validators on the BEVM chain verify the data provided by the Bitcoin light node client and reach consensus. Once consensus is achieved, the validated Bitcoin light node client data is synchronized into the EVM’s underlying account system. The Bitcoin asset holder can then use their Bitcoin on EVM to interact with Ethereum smart contracts and decentralized applications (DApps).

To transfer assets from the BEVM chain back to the Bitcoin blockchain, a cross-chain transaction must be submitted on the EVM platform. First, the $n$ PoS consensus nodes on BEVM perform a BFT vote using a BTC threshold custody contract, requiring more than two-thirds agreement. Once the vote passes, the threshold signature script is used as a leaf node in a MAST to generate a Taproot output key. This produces a Taproot transaction, which is then submitted to the Bitcoin blockchain, completing the cross-chain asset interaction. To spend this transaction, the threshold signature script must be used as an unlocking script to unlock the UTXO.

\subsection{STACKS}
\begin{figure}

  \centering
  \setlength{\abovecaptionskip}{0.cm}
  \includegraphics[width=\columnwidth]{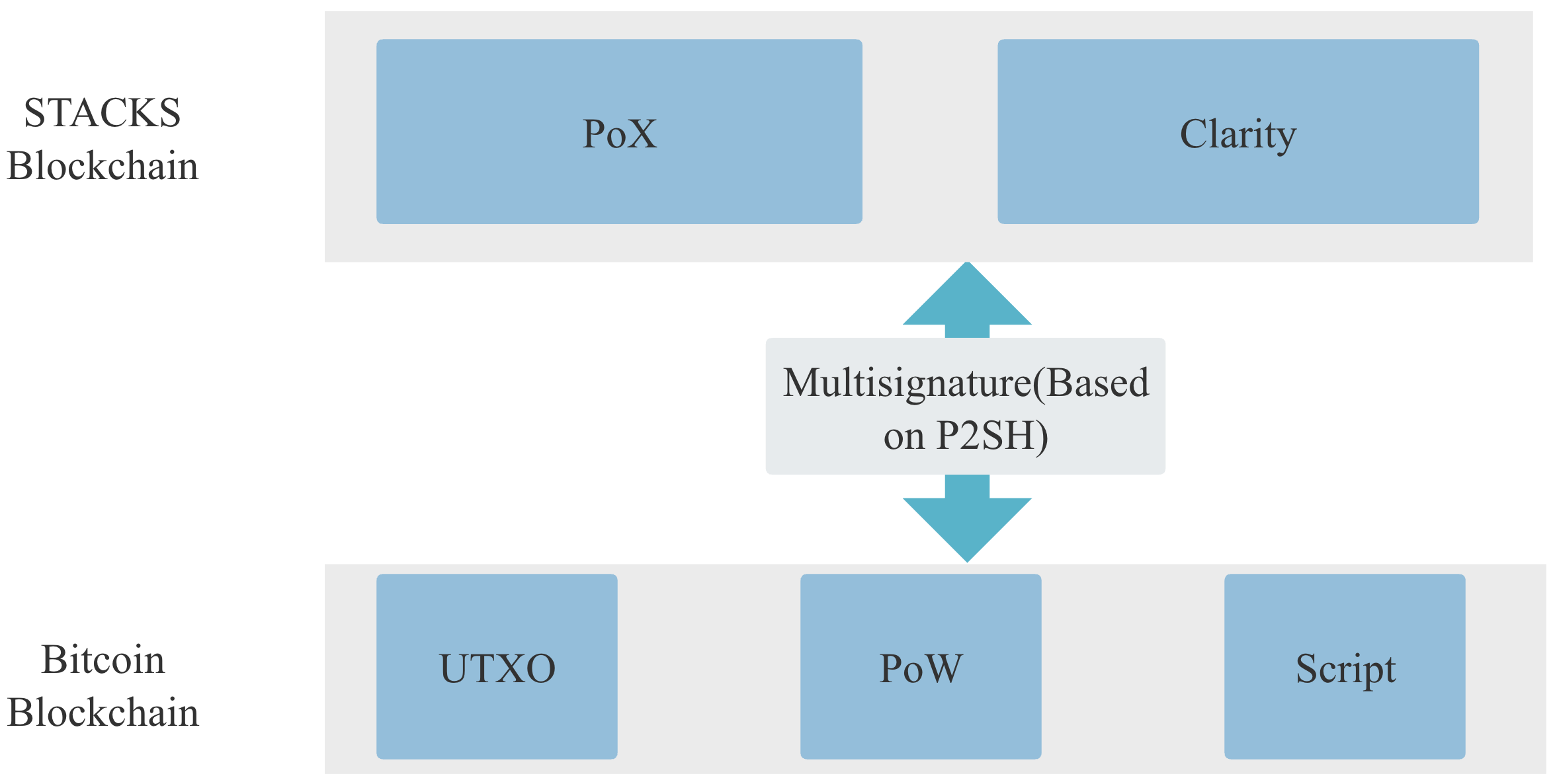}\\
  \caption{The architecture of STACKS.}\label{FIG. 20}

  \vspace{-0.5cm} \end{figure}

Stacks was launched in early 2021 as a smart contract layer for Bitcoin, using the Proof of Transfer (PoX) consensus protocol\cite{r14}. As shown in Fig. \ref{FIG. 20}, Stacks allows Bitcoin transaction settlement and uses a secure contract language called Clarity, which can respond to Bitcoin transactions. Additionally, Stacks supports atomic swaps between assets and BTC. The goal of Stacks is to transform Bitcoin from a passive asset into a productive one, supporting various decentralized applications to drive the growth of the Bitcoin economy. Similar to sidechains like RSK and Liquid, Stacks has its own global ledger and execution environment to support smart contracts without adding extra transaction burden to the Bitcoin blockchain. As shown in the figure, Stacks, like wBTC on Ethereum\cite{r27}, RBTC on RSK\cite{r24}, and L-BTC on Liquid\cite{r23}, achieves asset pegging by delegating the peg to centralized custodians or a trusted and permissioned federation of entities that manage Bitcoin peg transactions using multisignature methods.

Despite some progress in providing a smart contract layer for Bitcoin and introducing innovative features, Stacks' centralized peg mechanism faces several issues:
\begin{itemize}

\item [1.]Dependence on Centralized Entities: Stacks' centralized peg mechanism requires users to rely on centralized custodians or federated entities to manage the asset pegging and exchange process. This results in lower autonomy and control for users over their asset security and operations. Centralized entities may face security risks, misconduct, or single points of failure, posing potential threats to users' asset security.

\item [2.]Trust and Transparency Issues: The centralized peg mechanism requires users to trust that centralized entities will perform their duties correctly, with a lack of transparency and auditability. Users find it challenging to assess the fairness, compliance, and overall security of these entities' operations and decisions.

\item [3.]Security Risks: The centralized peg mechanism may face security risks. If centralized entities suffer from hacking, internal errors, or other security vulnerabilities, users' assets could be at risk of loss or theft. This poses a potential threat to asset security and may erode users' trust in the entire system.

\item [4.]Contradiction with Decentralization: Stacks' centralized peg mechanism contradicts the principle of decentralization. Although Stacks aims to provide more features and applications for Bitcoin, the centralized peg mechanism introduces centralized control and dependency.
\end{itemize}
Stacks' next major upgrade is the Nakamoto version, which primarily addresses the issue of centralized pegging of BTC assets in the original version by introducing a decentralized pegging solution that allows BTC to move in and out of the layer and be recorded on Bitcoin. Additionally, block production in Stacks will no longer be tied to miner elections, with miners producing blocks at a fixed rhythm. Since this upgrade requires a hard fork of the Stacks main chain, it needs approval from 70\% of stackers, making the upgrade plan as challenging to implement as reorganizing the Bitcoin mainchain.

\subsection{RGB++}

\begin{figure}

  \centering
  \setlength{\abovecaptionskip}{0.cm}
  \includegraphics[width=\columnwidth]{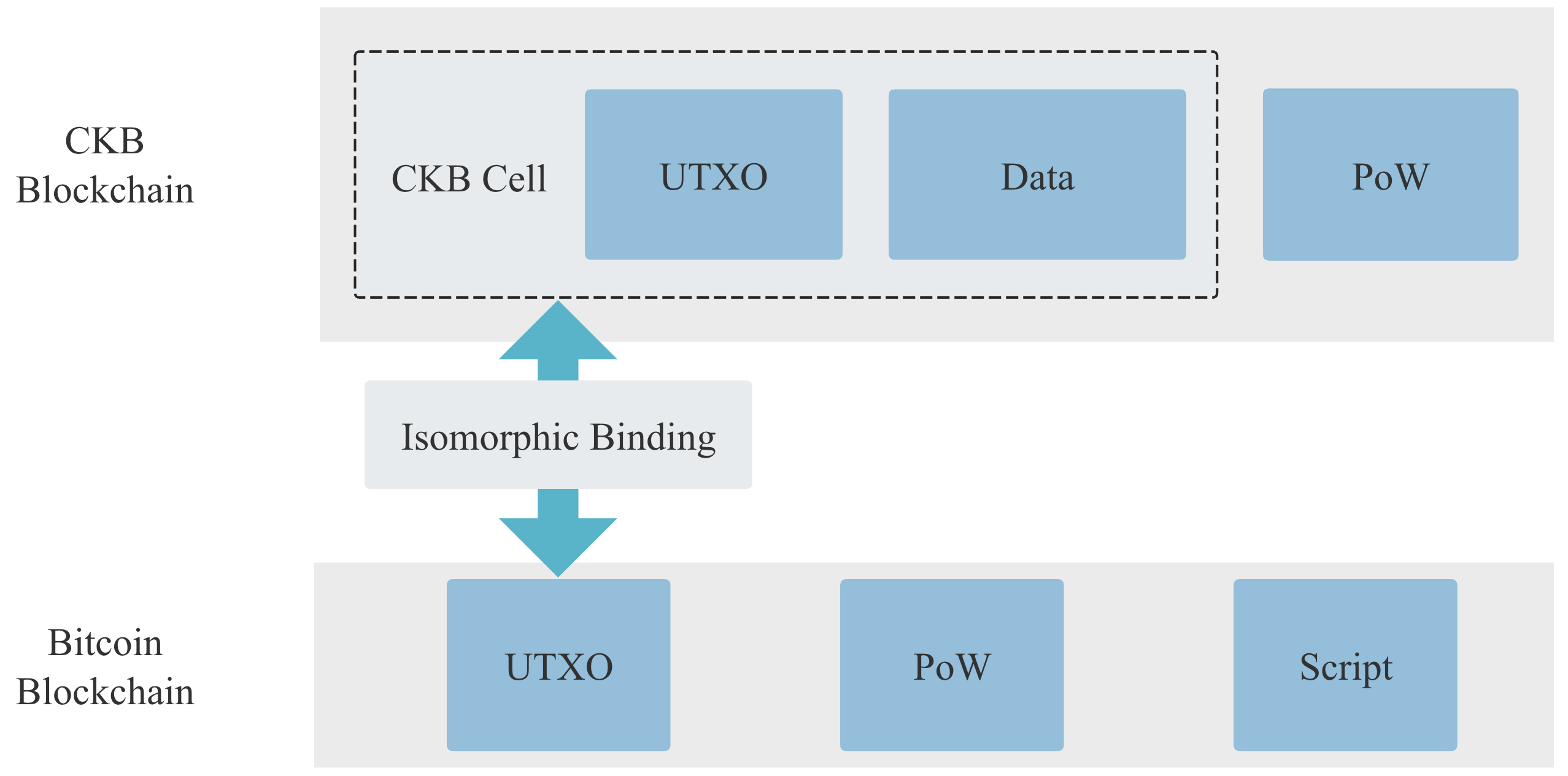}\\
  \caption{The architecture of RGB++.}\label{FIG. 21}

  \vspace{-0.5cm} \end{figure}

Nervos Network is an open-source blockchain ecosystem project that offers a set of solutions to address scalability and interoperability issues in blockchain. As shown in Fig. \ref{FIG. 21}, the core of the Nervos Network project is the CKB (Common Knowledge Base) chain, a public blockchain that uses a Proof of Work (PoW) consensus mechanism and an improved UTXO model for storing and securing encrypted assets. The improved UTXO model is called the Cell model. Similar to Bitcoin's UTXO, a Cell in the CKB chain is also a transaction output. However, unlike Bitcoin's UTXO, which only represents a certain number of sats and cannot store additional data beyond the locking script, a Cell on the CKB chain not only contains cryptocurrency value but also has storage capacity that can be used to store any data. In the CKB chain, $1$ CKB equals $1$ Byte of storage space. Therefore, if a Cell contains 99 CKB tokens, that Cell will have $99$ Bytes of on-chain storage space. This transforms the UTXO, which traditionally only stores an integer value representing token quantity, into a space capable of storing arbitrary data.

RGB++ is a protocol proposed by Cipher Wang, the co-founder of the Nervos Network project. This protocol provides a standard for issuing and managing assets on the Bitcoin blockchain. Notably, this protocol is integrated with the CKB chain as a Layer 2 solution for Bitcoin. Assets issued on the Bitcoin blockchain that comply with the RGB++ protocol (such as tokens, non-fungible tokens.) have their asset ledger managed on the CKB chain, where computation, verification, and data storage are handled. The RGB++ protocol, inspired by the RGB protocol, aims to enhance Bitcoin's performance by performing the computation, execution, and verification of token transactions off-chain, with final settlement occurring on the Bitcoin chain. Additionally, the RGB++ protocol binds asset ownership to Bitcoin's UTXO, ensuring that the settlement of new assets occurs on the Bitcoin chain, meaning that operations such as issuing, trading, and transferring new assets must be completed by spending a Bitcoin UTXO.

RGB++ introduces a concept called Isomorphic Binding, which differs from the form of cross-chain bridges. Isomorphic Binding functions by binding UTXOs on the Bitcoin blockchain that meet the RGB++ protocol to the CKB chain's Cell, using the Cell as a “container” for RGB asset data and writing the key parameters of RGB assets into the Cell. Since RGB assets are bound to Bitcoin UTXOs, the logical form of the assets retains the characteristics of UTXOs.

It is worth mentioning that Cipher Wang, as the leader of the CELL Studio team, along with his team, developed a modular Bitcoin Layer 2 issuance platform called UTXO STACK. The Bitcoin Layer 2 chain issued by this platform is isomorphic with Bitcoin's UTXO model.

\begin{table*}[!t]

\caption{Comparison of Layer2 Protocols}\label{Table II}

\centering
\begin{tabular}{c|ccccc}
Protocol & Category & Smart contract & Consensus protocol & Bridging Technology & Language \\
\hline
          BitVM & Rollup & Yes & - & - & Bitcoin script \\
          Babylon & Sidechain & Yes & PoS & - & Go \\
          BEVM & Sidechain & Yes & PoS & Multisignature(Based on P2TR) & Solidity \\
          STACKS & Sidechain & Yes & PoX & Multisignature(Based on P2SH) & Clarity \\
          RGB++ & Sidechain & Yes & PoW & Isomorphic Binding & C \\
          RGB & Client-side verification & Yes & - & - & Contractum \\
          Lightning Network & State Channel & Not & - & - & - \\
\end{tabular}
\end{table*}

\subsection{RGB}
The RGB protocol, developed by the LNP/BP Standards Association, is a scalable and privacy-focused smart contract system for Bitcoin and the Lightning Network\cite{r31}\cite{r29}. The core technological components of RGB were proposed by Peter Todd in 2017, consisting of two main concepts: single-use-seals and client-side validation\cite{r25}.

In RGB, tokens are associated with a Bitcoin UTXO (whether an existing UTXO or a temporarily created one). To transfer these tokens, the associated UTXO must be spent. When spending this UTXO, the Bitcoin transaction must include a commitment to a message containing the RGB payment information. This message defines the inputs, the UTXO to which the tokens will be sent, the asset ID, the amount, the spending transaction, and any other necessary data. Essentially, the core data running on RGB attaches itself to the Bitcoin blockchain via UTXO, leveraging the Bitcoin mainnet to secure the assets. However, this functionality faces two challenges:
\begin{itemize}

  \item [1.] Verification Complexity: Since the client-side validation of assets requires tracing every upstream UTXO of the asset, a significant amount of data verification is involved. The more times an asset is transferred, the greater the difficulty and cost of validation.

  \item [2.] Off-Chain Data Dependence: Although assets can be verified, the Bitcoin blockchain only serves as a ledger for off-chain data. Bitcoin miners do not actually participate in the validation of RGB assets, necessitating third-party validation of the RGB ledger. Additionally, RGB smart contracts do not truly operate on-chain; each RGB-based smart contract is independent and cannot interact with others. For instance, if two tokens issued on RGB need to be swapped, this cannot be done directly like assets issued on the Ethereum Virtual Machine (EVM). Instead, the tokens must be transferred to the Lightning Network for interaction.
\end{itemize}

\subsection{Lightning Network}
We classify the Lightning Network as a state channel Bitcoin Layer 2 scaling solution. The Lightning Network was first proposed in a white paper written by Joseph Poon and Tadge Dryja in 2015. Its core idea is to use payment channels to avoid frequent interactions with the blockchain. Specifically, the Lightning Network allows participants to establish a two-way payment channel on the blockchain. Once the channel is established, participants can conduct multiple fast offline transactions within the channel. These transactions do not need to be recorded directly on the blockchain. Only when the final settlement is completed, the participants submit the transaction results to the blockchain. This greatly reduces the load and transaction costs of the Bitcoin. The emergence of the Lightning Network solves the scalability problem of the Bitcoin blockchain. Due to the slow creation speed of Bitcoin blocks and the limited number of transactions accommodated, transaction fees are high and transaction times are long. The use of the Lightning Network can improve the efficiency of block space usage, making small payments and high-frequency transactions more convenient and practical. The Lightning Network also has a high degree of privacy protection. The transaction information under the user chain will not spread on the network, and the transaction details are only visible to the parties involved. The underlying principles of the Lightning Network include multisignature addresses and hash time lock contracts (HTLC). Multisignature addresses allow users to specify the number of private keys required to pay funds and sign transactions. In the Lightning Network channel, participants lock funds through the $2$-of-$2$ signature scheme, and two private keys are also required to transfer funds. A hashed timelock contract is a mechanism that enforces a contract between participants, allowing the other party to take remedial action and withdraw funds if one party does not play by the rules.

\subsection{Bitcoin Layer 2 Protocol Comparison}
We presents a comparison of several representative Bitcoin Layer 2 protocols, focusing on their functionalities, architectures, and implementations. The comparison results are summarized in TABLE \ref{Table II}, which highlights key aspects such as protocol category, smart contract support, consensus mechanism, the technology of bridging, and programming language.

To begin with, in terms of protocol categories, BitVM stands as a representative of Rollup technology, combining off-chain computation with on-chain validation to significantly enhance transaction processing capacity. This makes it an important solution for addressing Bitcoin's scalability challenges. In contrast, protocols such as Babylon, BEVM, STACKS, and RGB++ are sidechain solutions that operate parallel to the Bitcoin mainchain, providing enhanced scalability while enriching the functionality and diversity of the Bitcoin ecosystem. Notably, the Lightning Network, as a representative of state channel technology, takes a different approach by leveraging existing blockchain resources without creating a new blockchain, streamlining transaction processes to achieve highly efficient payment handling.

The ability for smart contracts is a crucial factor in assessing the potential applications of Layer 2 protocols. BitVM, through its innovative Rollup architecture, extends Bitcoin's functionality to support complex smart contracts. Babylon and BEVM, as sidechain protocols, embed smart contract support directly into their respective blockchain environments. Meanwhile, RGB utilizes the underlying architecture of the Lightning Network to enable indirect support for smart contracts.

Consensus mechanisms play a critical role in determining the level of decentralization and security of Bitcoin Layer 2 sidechains. Babylon and BEVM adopt Proof of Stake (PoS), encouraging active participation from network nodes through economic incentives, which ensures a high degree of decentralization. STACKS introduces the innovative Proof of Transfer (PoX) consensus, which has proven effective in maintaining network security and decentralization. RGB++, on the other hand, employs a Proof of Work (PoW) consensus, using complex mathematical problems to secure the network and guarantee its immutability.

When evaluating the security of sidechain protocols, the technology of bridging with the Bitcoin blockchain are crucial. The STACKS protocol employs a more traditional P2SH-based multisignature scheme to achieve cross-chain interactions with the Bitcoin blockchain. While this scheme effectively meets cross-chain requirements, it has significant limitations, as its multisignature mechanism supports only up to 15 participants. This limitation hinders broader node participation and reduces the decentralization of decision-making, which could impact the overall security and decentralization of the sidechain protocol. In contrast, the BEVM protocol demonstrates notable innovation in this regard by adopting a P2TR-based multisignature scheme as its bridging technology. This approach supports a larger number of participants (up to 1,000 participants within the protocol) to collaborate in the multisignature process. This significantly increases the participation breadth of nodes and the democratization of decision-making, thereby enhancing the security and decentralization of the network.

On the other hand, RGB++ employs a unique technology called Isomorphic Binding, which directly binds UTXOs from the Bitcoin blockchain, representing cross-chain assets in an alternative form. While technically distinct, this method focuses more on UTXO-level binding than on enhancing the security and decentralization of interactions through multisignature mechanisms.

\section{Data Collection and Analysis}
\label{sec:Data Collection and Analysis}
In order to scientifically evaluate the impact and actual results of the upgrade of Taproot proposal and its derived protocols (such as Ordinals, Atomicals) on the Bitcoin mainnet, we conducted data collection and analysis on the transactions on the Bitcoin blockchain. Specifically, we crawled all transactions on the Bitcoin blockchain from January 3, 2017, to August 21, 2024, and analyzed these transaction data to obtain the changes in the average block size per day, the usage trend of the P2TR transaction type, and the fluctuation of miner fees on the Bitcoin blockchain caused by Taproot-like transaction activities (especially those initiated by protocols such as Ordinals, Atomicals and Runes).

\begin{figure}[!t]

  \centering
  \setlength{\abovecaptionskip}{0.cm}
  \includegraphics[width=\columnwidth]{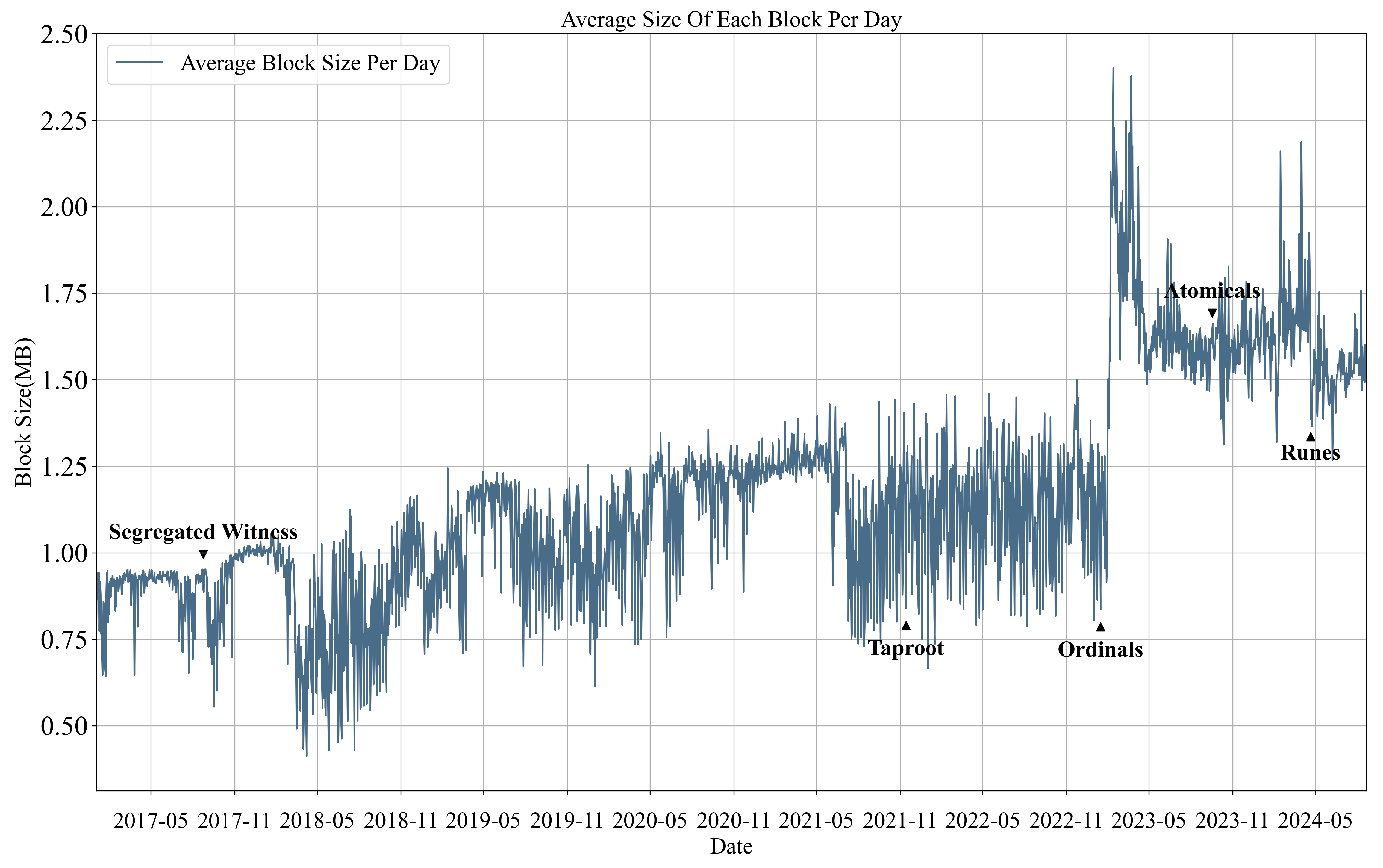}\\
  \caption{The average size of each block per day on the Bitcoin blockchain.}\label{FIG. 22}

  \vspace{-0.1cm} \end{figure}

Fig. \ref{FIG. 22} depicts the trend of the average block size per day on the Bitcoin blockchain from January 2017 to August 2024. It can be observed from the figure that after the two technical upgrades of 'Segregated Witness' and 'Taproot proposal' (on August 24, 2017, and November 14, 2021, respectively), the average block size per day showed an overall upward trend. Several factors contributed to this phenomenon. Firstly, key technological upgrades such as SegWit and Taproot optimized transaction data structures, enhanced blockchain scalability, and reduced data redundancy, significantly increasing the effective capacity of transactions within blocks, allowing more transactions to be accommodated within the 1MB physical block size limit. Secondly, the introduction of new protocols based on Taproot (Ordinals, Atomicals, Runes, etc.) not only enriched the functionality and application scenarios of the Bitcoin mainnet but also had an impact on the block space requirements by introducing new data structures. These new protocols often involve more data elements or require more complex script execution, resulting in a larger block space occupied by a single transaction compared to traditional transactions. With the block size limit remaining the same, this means that when accommodating the same number of traditional transactions (measured by transaction count), more space is needed to store them due to the presence of complex transactions.

\begin{figure}[!t]

  \centering
  \setlength{\abovecaptionskip}{0.cm}
  \includegraphics[width=\columnwidth]{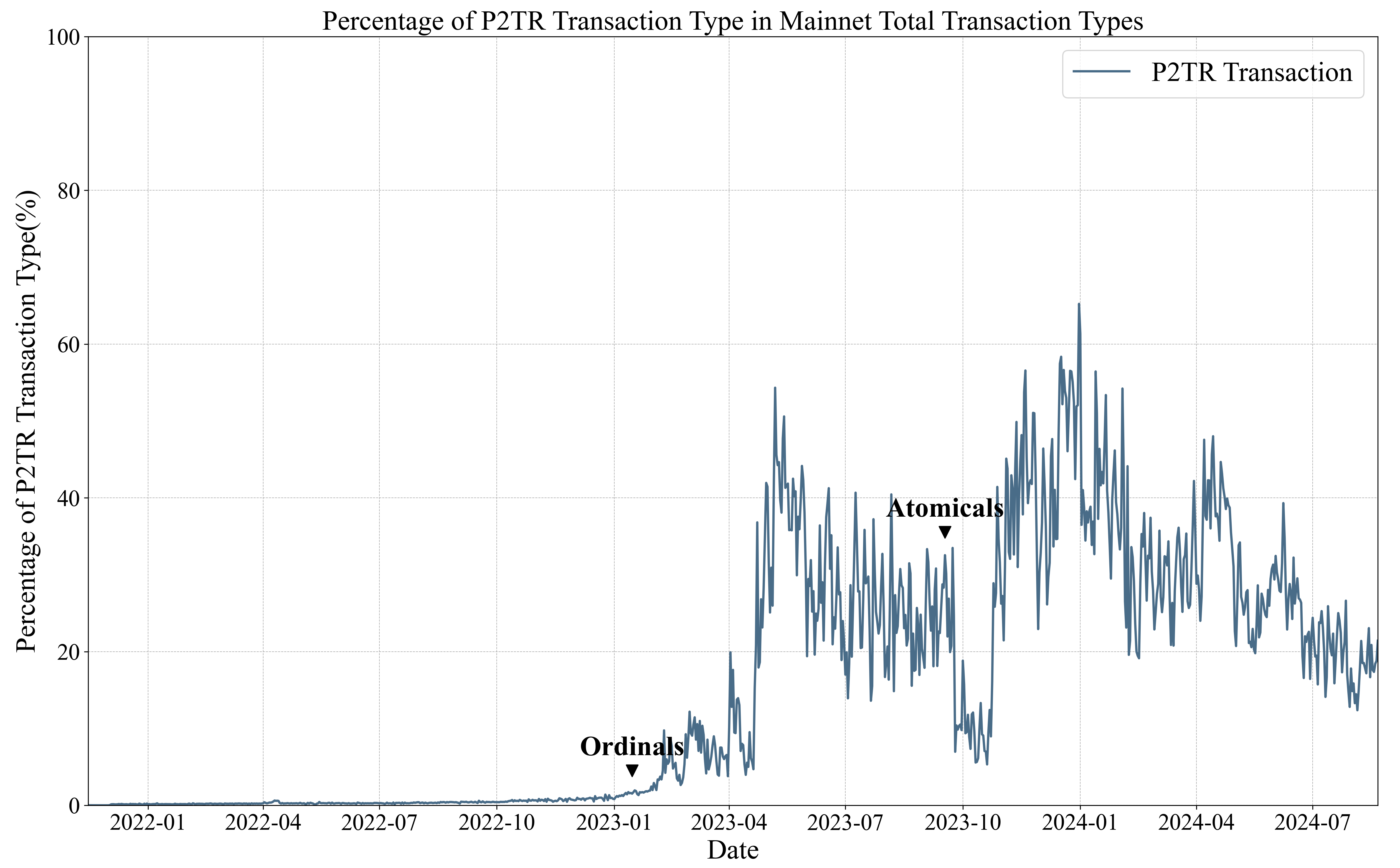}\\
  \caption{The percentage of P2TR transaction type on the Bitcoin blockchain.}\label{FIG. 23}
  \vspace{-0.1cm} \end{figure}

Fig. \ref{FIG. 23} depicts the change in the percentage of P2TR transaction type in the total transaction types on the Bitcoin blockchain from January 2023 to August 2024. It can be seen from the figure that with the emergence of protocols such as Ordinals and Atomicals, P2TR transactions gradually occupy an important position among the various transaction types on the Bitcoin blockchain (between approximately 20\% and 50\%). As mentioned earlier, the deploy, mint, and transfer operations of NFTs and fungible tokens in protocols such as Ordinals and Atomicals rely on the P2TR transaction type. The data in Fig. \ref{FIG. 23} indicates the successful application of protocols such as Ordinals and Atomicals, which indeed greatly boost the proportion of P2TR transactions on the Bitcoin blockchain.

\begin{figure}[!t]

  \centering
  \setlength{\abovecaptionskip}{0.cm}
  \includegraphics[width=\columnwidth]{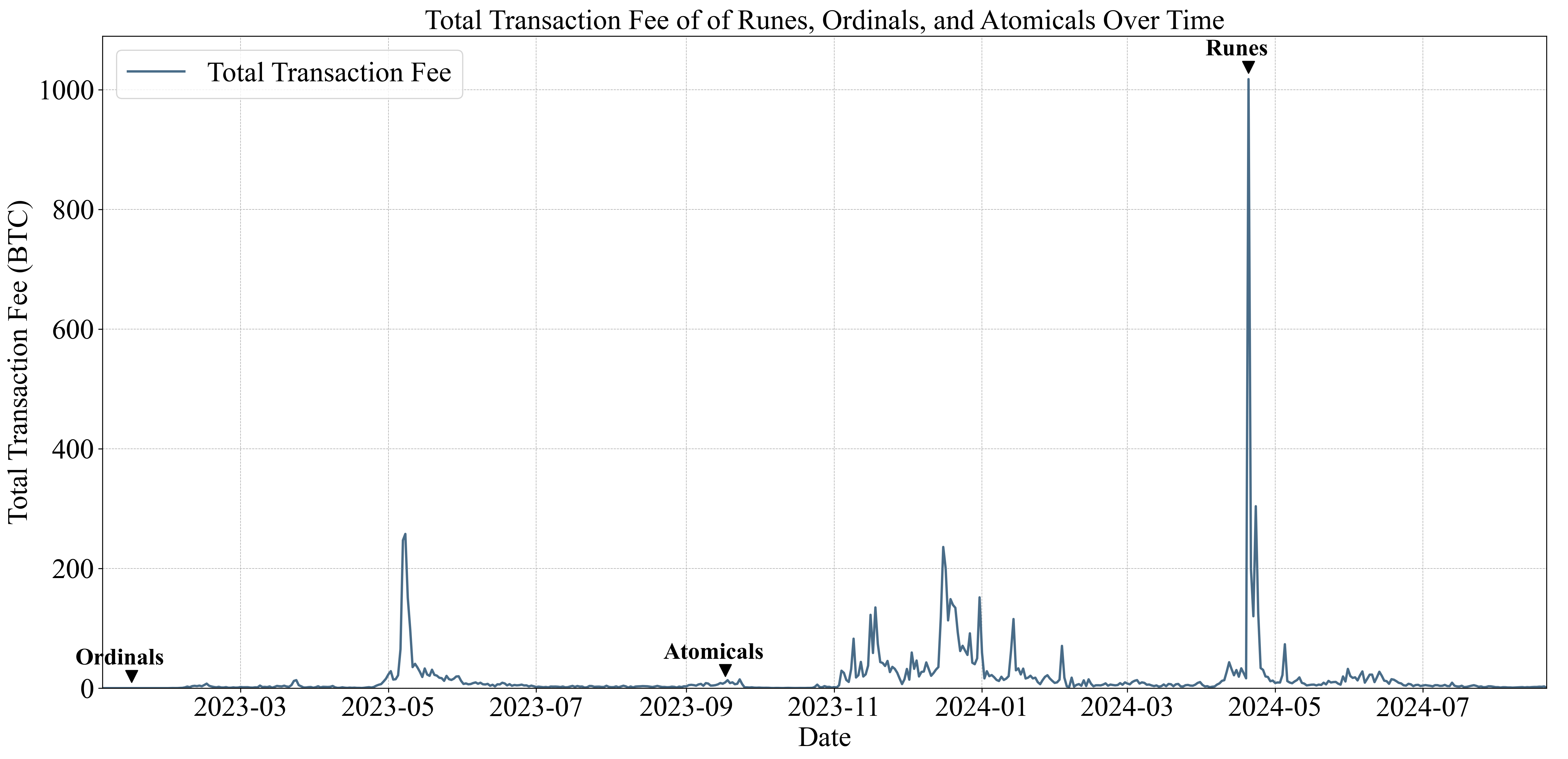}\\
  \caption{The transaction fees used by the BRC-20, ARC-20 and Runes operations.}\label{FIG. 24}

  \vspace{-0.5cm} \end{figure}

Fig. \ref{FIG. 24} depicts the transaction fees required for the deploy, mint, and transfer operations of fungible tokens in protocols such as BRC-20, ARC-20, and Runes on the Bitcoin blockchain from January 2023 to July 2024. It can be observed from the figure that these fungible token standards in Bitcoin have attracted significant market attention, leading to a large number of users participating in token deploy, mint, and transfer operations, thereby driving up transaction fees on the Bitcoin blockchain. The fees generated from these token operations become a substantial source of income for miners.

\section{Conclusion}\label{conclusion}

This comprehensive survey has elucidated the significant advancements in Bitcoin's Layer 1 and Layer 2 protocols, which have substantially enhanced its programming functionalities. The upgrade brought by the Taproot proposal in 2021, introducing the Schnorr signature algorithm and P2TR transaction type, has catalyzed remarkable improvements in Bitcoin's privacy and programmability. This evolution has fostered the development of innovative protocols such as Ordinals, Atomicals, and BitVM, expanding Bitcoin's applications into realms previously thought impractical, including non-fungible tokens (NFTs) and more sophisticated fungible token standards (e.g., BRC-20, ARC-20). Our analysis of transaction data from the Bitcoin blockchain revealed positive trends, including increased block capacity and miner revenues driven by these new protocols. The findings affirm that these Layer 2 solutions not only address the scalability challenges that have historically impeded Bitcoin's growth but also open up new avenues for application development within the network. In conclusion, this survey provides a critical foundation for understanding Bitcoin's technological trajectory. We illuminate how Layer 1 and Layer 2 innovations are synergistically enhancing Bitcoin's capabilities, potentially catalyzing broader adoption and diversifying its applications. 

Despite these advancements, several limitations remain. The integration of complex functionalities into Bitcoin's blockchain can lead to increased transaction sizes and potential network congestion. Furthermore, reliance on off-chain solutions like Lightning Network requires robust security measures to ensure the integrity of transactions processed outside the mainchain. Future research should focus on optimizing scalability solutions to handle increased transaction volumes without compromising speed or security. Exploring interoperability between Bitcoin's ecosystem and other blockchain platforms could also open new avenues for cross-chain applications. Additionally, developing more efficient privacy-preserving mechanisms will be crucial as Bitcoin continues to expand its capabilities. Overall, while the current protocols have enhanced Bitcoin's ecosystem, ongoing innovation and research are essential to address existing challenges and unlock further potential in decentralized finance and beyond.


%





\ifCLASSOPTIONcaptionsoff
  \newpage
\fi




\bibliographystyle{IEEEtran}
 \bibliography{main} 

\begin{thebibliography}{10}
\providecommand{\url}[1]{#1}
\csname url@rmstyle\endcsname
\providecommand{\newblock}{\relax}
\providecommand{\bibinfo}[2]{#2}
\providecommand\BIBentrySTDinterwordspacing{\spaceskip=0pt\relax}
\providecommand\BIBentryALTinterwordstretchfactor{4}
\providecommand\BIBentryALTinterwordspacing{\spaceskip=\fontdimen2\font plus
\BIBentryALTinterwordstretchfactor\fontdimen3\font minus \fontdimen4\font\relax}
\providecommand\BIBforeignlanguage[2]{{%
\expandafter\ifx\csname l@#1\endcsname\relax
\typeout{** WARNING: IEEEtran.bst: No hyphenation pattern has been}%
\typeout{** loaded for the language `#1'. Using the pattern for}%
\typeout{** the default language instead.}%
\else
\language=\csname l@#1\endcsname
\fi
#2}}

\bibitem{r10}
S.~Nakamoto, ``Bitcoin: {{A Peer-to-Peer Electronic Cash System}},'' p.~9.

\bibitem{r48}
J.~Lovejoy, M.~Virza, C.~Fields, K.~Karwaski, A.~Brownworth, and N.~Narula, ``Hamilton: A $\{$High-Performance$\}$ transaction processor for central bank digital currencies,'' in \emph{20th USENIX Symposium on Networked Systems Design and Implementation (NSDI 23)}, 2023, pp. 901--915.

\bibitem{r7}
A.~M. Antonopoulos, \emph{Mastering Bitcoin: Unlocking Digital Cryptocurrencies}, 1st~ed.\hskip 1em plus 0.5em minus 0.4em\relax Beijing K{\"o}ln: O'Reilly, 2015.

\bibitem{r47}
M.~Bartoletti and L.~Pompianu, ``An analysis of bitcoin op{\_}return metadata,'' in \emph{Financial Cryptography and Data Security}.\hskip 1em plus 0.5em minus 0.4em\relax Cham: Springer International Publishing, 2017, pp. 218--230.

\bibitem{r16}
\BIBentryALTinterwordspacing
``Bips/bip-0114.mediawiki at master {$\cdot$} bitcoin/bips.'' [Online]. Available: \url{https://github.com/bitcoin/bips/blob/master/bip-0114.mediawiki}
\BIBentrySTDinterwordspacing

\bibitem{r12}
J.~Poon and T.~Dryja, ``The {{Bitcoin Lightning Network}}:.''

\bibitem{r17}
\BIBentryALTinterwordspacing
``Bips/bip-0141.mediawiki at master {$\cdot$} bitcoin/bips.'' [Online]. Available: \url{https://github.com/bitcoin/bips/blob/master/bip-0141.mediawiki\#user-content-Abstract}
\BIBentrySTDinterwordspacing

\bibitem{r18}
\BIBentryALTinterwordspacing
``Bips/bip-0340.mediawiki at master {$\cdot$} bitcoin/bips.'' [Online]. Available: \url{https://github.com/bitcoin/bips/blob/master/bip-0340.mediawiki}
\BIBentrySTDinterwordspacing

\bibitem{r19}
\BIBentryALTinterwordspacing
``Bips/bip-0341.mediawiki at master {$\cdot$} bitcoin/bips.'' [Online]. Available: \url{https://github.com/bitcoin/bips/blob/master/bip-0341.mediawiki}
\BIBentrySTDinterwordspacing

\bibitem{r4}
A.~Jain and E.~S. Pilli, ``{{SoK}}: {{Digital Signatures}} and~{{Taproot Transactions}} in~{{Bitcoin}},'' in \emph{Information {{Systems Security}}}, ser. Lecture {{Notes}} in {{Computer Science}}, V.~Muthukkumarasamy, S.~D. Sudarsan, and R.~K. Shyamasundar, Eds.\hskip 1em plus 0.5em minus 0.4em\relax Cham: Springer Nature Switzerland, 2023, pp. 360--379.

\bibitem{r45}
N.~Li, M.~Qi, Q.~Wang, and S.~Chen, ``Bitcoin inscriptions: Foundations and beyond,'' \emph{arXiv preprint arXiv:2401.17581}, 2024.

\bibitem{r2}
S.~{Delgado-Segura}, C.~{P{\'e}rez-Sol{\`a}}, G.~{Navarro-Arribas}, and J.~{Herrera-Joancomart{\'i}}, ``Analysis of the {{Bitcoin UTXO Set}},'' in \emph{Financial {{Cryptography}} and {{Data Security}}}, ser. Lecture {{Notes}} in {{Computer Science}}, A.~Zohar, I.~Eyal, V.~Teague, J.~Clark, A.~Bracciali, F.~Pintore, and M.~Sala, Eds.\hskip 1em plus 0.5em minus 0.4em\relax Berlin, Heidelberg: Springer, 2019, pp. 78--91.

\bibitem{r11}
K.~Okupski, ``Bitcoin {{Developer Reference}}.''

\bibitem{r3}
B.~Hou and F.~Chen, ``A {{Study}} on {{Nine Years}} of {{Bitcoin Transactions}}: {{Understanding Real-world Behaviors}} of {{Bitcoin Miners}} and {{Users}},'' in \emph{2020 {{IEEE}} 40th {{International Conference}} on {{Distributed Computing Systems}} ({{ICDCS}})}, Nov. 2020, pp. 1031--1043.

\bibitem{r20}
\BIBentryALTinterwordspacing
``Bitcoinbook/ch07 authorization-authentication.'' [Online]. Available: \url{https://github.com/bitcoinbook/bitcoinbook/blob/develop/ch07\_authorization-authentication.adoc}
\BIBentrySTDinterwordspacing

\bibitem{r42}
\BIBentryALTinterwordspacing
``Bips/bip-0032.mediawiki at master {$\cdot$} bitcoin/bips.'' [Online]. Available: \url{https://github.com/bitcoin/bips/blob/master/bip-0032.mediawiki}
\BIBentrySTDinterwordspacing

\bibitem{r43}
G.~Neven, N.~P. Smart, and B.~Warinschi, ``Hash function requirements for schnorr signatures,'' \emph{Journal of Mathematical Cryptology}, vol.~3, no.~1, pp. 69--87, 2009.

\bibitem{r5}
F.~Kleinwort, W.~Posdorfer, and J.~Edinger, ``Analyzing the effect of taproot on bitcoin deanonymization,'' in \emph{2023 IEEE 43rd International Conference on Distributed Computing Systems Workshops ICDCSW}, July 2023, pp. 25--30.

\bibitem{r6}
J.~Xie, F.~R. Yu, T.~Huang, R.~Xie, J.~Liu, and Y.~Liu, ``A {{Survey}} on the {{Scalability}} of {{Blockchain Systems}},'' \emph{IEEE Network}, vol.~33, no.~5, pp. 166--173, Sept. 2019.

\bibitem{r41}
J.~Rubin and M.~Naik, ``Merkelized {{Abstract Syntax Trees}},'' 2014.

\bibitem{r44}
\BIBentryALTinterwordspacing
``Bips/bip-0342.mediawiki at master {$\cdot$} bitcoin/bips.'' [Online]. Available: \url{https://github.com/bitcoin/bips/blob/master/bip-0342.mediawiki}
\BIBentrySTDinterwordspacing

\bibitem{r26}
\BIBentryALTinterwordspacing
``Simple schnorr multi-signatures with applications to bitcoin designs, codes and cryptography.'' [Online]. Available: \url{https://link.springer.com/article/10.1007/s10623-019-00608-x}
\BIBentrySTDinterwordspacing

\bibitem{r34}
E.~Crites, C.~Komlo, and M.~Maller, ``How to {{Prove Schnorr Assuming Schnorr}}: {{Security}} of {{Multi-}} and {{Threshold Signatures}},'' 2021.

\bibitem{r22}
\BIBentryALTinterwordspacing
``Introduction - {{Ordinal Theory Handbook}}.'' [Online]. Available: \url{https://docs.ordinals.com/}
\BIBentrySTDinterwordspacing

\bibitem{r30}
\BIBentryALTinterwordspacing
``Introduction {\textbar} {{Atomicals Guidebook}},'' Jan. 2024. [Online]. Available: \url{https://docs.atomicals.xyz}
\BIBentrySTDinterwordspacing

\bibitem{sguanci2021layer}
C.~Sguanci, R.~Spatafora, and A.~M. Vergani, ``Layer 2 blockchain scaling: A survey,'' \emph{arXiv preprint arXiv:2107.10881}, 2021.

\bibitem{r9}
R.~Linus, ``{{BitVM}}: {{Compute Anything}} on {{Bitcoin}}.''

\bibitem{r8}
M.~Armstrong, ``Ethereum, {{Smart Contracts}} and the {{Optimistic Roll-up}}.''

\bibitem{r35}
\BIBentryALTinterwordspacing
``{REASONING ABOUT DIGITAL CIRCUITS - ProQuest}.'' [Online]. Available: \url{https://www.proquest.com/openview}
\BIBentrySTDinterwordspacing

\bibitem{r39}
\BIBentryALTinterwordspacing
``{{BabylonChain}}.'' [Online]. Available: \url{https://babylonchain.io/}
\BIBentrySTDinterwordspacing

\bibitem{r38}
E.~N. Tas, D.~Tse, F.~Gai, S.~Kannan, M.~A. {Maddah-Ali}, and F.~Yu, ``Bitcoin-{{Enhanced Proof-of-Stake Security}}: {{Possibilities}} and {{Impossibilities}},'' May 2023.

\bibitem{r13}
F.~Saleh, ``Blockchain {{Without Waste}}: {{Proof-of-Stake}}.''

\bibitem{r40}
\BIBentryALTinterwordspacing
``Github - osmosis-labs/mesh-security.'' [Online]. Available: \url{https://github.com/osmosis-labs/mesh-security?tab=readme-ov-file}
\BIBentrySTDinterwordspacing

\bibitem{r37}
\BIBentryALTinterwordspacing
``Btc\_staking\_litepaper({{EN}}).pdf.'' [Online]. Available: \url{https://docs.babylonchain.io/papers/btc\_staking\_litepaper(EN).pdf}
\BIBentrySTDinterwordspacing

\bibitem{shi2022pooling}
L.~Shi, T.~Wang, J.~Li, S.~Zhang, and S.~Guo, ``Pooling is not favorable: Decentralize mining power of pow blockchain using age-of-work,'' \emph{IEEE Transactions on Cloud Computing}, vol.~11, no.~3, pp. 2756--2769, 2022.

\bibitem{r36}
E.~Deirmentzoglou, G.~Papakyriakopoulos, and C.~Patsakis, ``A {{Survey}} on {{Long-Range Attacks}} for {{Proof}} of {{Stake Protocols}},'' \emph{IEEE Access}, vol.~7, pp. 28\,712--28\,725, 2019.

\bibitem{r21}
\BIBentryALTinterwordspacing
``{{GitHub}} - btclayer2/{{BEVM-white-paper}}.'' [Online]. Available: \url{https://github.com/btclayer2/BEVM-white-paper/tree/main}
\BIBentrySTDinterwordspacing

\bibitem{r32}
``Ethereum/tests,'' ethereum, Mar. 2024.

\bibitem{r15}
\BIBentryALTinterwordspacing
A.~Stewart and E.~{Kokoris-Kogia}, ``{{GRANDPA}}: A {{Byzantine Finality Gadget}},'' July 2020. [Online]. Available: \url{https://arxiv.org/abs/2007.01560v1}
\BIBentrySTDinterwordspacing

\bibitem{r14}
``Stacks: {{A Bitcoin Layer}} for {{Smart Contracts}}.''

\bibitem{r27}
\BIBentryALTinterwordspacing
``{{WBTC Wrapped Bitcoin}} an {{ERC20}} token backed 1:1 with {{Bitcoin}}.'' [Online]. Available: \url{https://www.wbtc.network}
\BIBentrySTDinterwordspacing

\bibitem{r24}
\BIBentryALTinterwordspacing
``{{RBTC}}: Smart bitcoin powering the {{RSK}} network {\textbar} {{Rootstock}} ({{RSK}}).'' [Online]. Available: \url{https://rootstock.io/rbtc/}
\BIBentrySTDinterwordspacing

\bibitem{r23}
\BIBentryALTinterwordspacing
``Liquid {{Network}}: {{Purpose-Built}} for {{Asset Issuance}}.'' [Online]. Available: \url{https://liquid.net}
\BIBentrySTDinterwordspacing

\bibitem{r31}
\BIBentryALTinterwordspacing
``What is {{RGB}}? {\textbar} {{RGB FAQ}},'' Feb. 2024. [Online]. Available: \url{https://www.rgbfaq.com/what-is-rgb}
\BIBentrySTDinterwordspacing

\bibitem{r29}
\BIBentryALTinterwordspacing
``{{RGB Blackpaper}} {\textbar} {{RGB Blackpaper}},'' Mar. 2023. [Online]. Available: \url{https://blackpaper.rgb.tech}
\BIBentrySTDinterwordspacing

\bibitem{r25}
\BIBentryALTinterwordspacing
``Scalable {{Semi-Trustless Asset Transfer}} via {{Single-Use-Seals}} and {{Proof-of-Publication}}.'' [Online]. Available: \url{https://petertodd.org/2017/scalable-single-use-seal-asset-transfer}
\BIBentrySTDinterwordspacing

\end{thebibliography}

\vfill


\end{document}